\documentclass[pdflatex,sn-mathphys-num]{sn-jnl}

\usepackage{graphicx}%
\usepackage{multirow}%
\usepackage{amsmath,amssymb,amsfonts}%
\usepackage{amsthm}%
\usepackage{mathrsfs}%
\usepackage[title]{appendix}%
\usepackage{xcolor}%
\usepackage{textcomp}%
\usepackage{manyfoot}%
\usepackage{booktabs}%
\usepackage{algorithm}%
\usepackage{algorithmicx}%
\usepackage{algpseudocode}%
\usepackage{listings}%

\usepackage[bitstream-charter]{mathdesign}
\usepackage{amsmath}
\usepackage{amsthm}
\usepackage{mathrsfs} 

\usepackage[
  math-style=ISO,
  bold-style=ISO,
  sans-style=italic,
  nabla=upright,
  partial=upright,
  warnings-off={mathtools-colon,mathtools-overbracket}, 
]{unicode-math}
\setoperatorfont\symup

\usepackage{physics}
\usepackage{braket}
\usepackage{mathtools}
\usepackage{cancel} 
\usepackage[
  separate-uncertainty=true,
  per-mode=symbol-or-fraction,
]{siunitx}

\usepackage{epsfig}
\usepackage{epic}
\usepackage{eepic}
\usepackage{amscd}
\usepackage{graphicx}
\usepackage{pgfplots}

\usepackage{threeparttable} 
\usepackage{tabularx} 
\usepackage{longtable} 
\usepackage{dcolumn} 
\usepackage{booktabs}  

\usepackage{caption}
\usepackage{subcaption}
\usepackage{here} 
\usepackage{placeins} 

\usepackage{lscape}
\usepackage{color}
\usepackage{blindtext} 
\usepackage{changepage} 
\usepackage{dsfont}
\usepackage{ifthen} 

\usepackage[capitalise]{cleveref}
\crefname{section}{Sec.}{sections}
\crefname{appendix}{App.}{appendizes}

\usepackage[
    acronym,
    hyperfirst = false
]{glossaries}
\glsdisablehyper 
\newacronym{cst}{CST}{continuous similarity transformation}
\newacronym{cut}{CUT}{continuous unitary transformation}
\newacronym{sct}{scSWT}{self-consistent spin wave theory}
\newacronym{lswt}{LSWT}{linear spin-wave theory}
\newacronym{qpc}{qpc}{quasi-particle conserving}
\newacronym{ins}{INS}{inelastic neutron scattering}
\newacronym{pbc}{pbc}{periodic boundary conditions}
\newacronym{apbc}{apbc}{antiperiodic boundary conditions}
\newacronym{bz}{BZ}{Brillouin zone}
\newacronym{rod}{ROD}{residual off-diagonality}
\newacronym{nnn}{NNN}{next-nearest neighbor}
\newacronym{nn}{NN}{nearest neighbor}

\newcommand{\neel}{N\'eel}
\newcommand{\deltaSpin}{\ensuremath{\Delta S}}

\newcommand{\dsc}{d_\text{sc}}

 \NewDocumentCommand\Vector{g} {\ensuremath{  
  \symbf{#1}
} }
\NewDocumentCommand\Matrix{g} {\ensuremath{  \underline{\bm{#1}}  } }
\NewDocumentCommand\Operator{m}{\ensuremath{ \hat{#1} } }

\AtBeginDocument{\renewcommand{\Im}{\operatorname{Im}}}

\NewDocumentCommand\kvec{m}{\ensuremath{\Vector{k}_{#1}}}

\NewDocumentCommand\cexpval{g}{\ensuremath{  \Braket{#1}}_0 }

\NewDocumentCommand\norord{g}{\ensuremath{:\!{#1}\!:}}

\NewDocumentCommand\abos{ s o }{ \ensuremath{ \IfBooleanTF#1
{ \hat{a}^{\dagger}_{#2}    } { \hat{a}^{\phantom{\dagger}}_{#2} } } }

\NewDocumentCommand\bbos{ s o }{ \ensuremath{ \IfBooleanTF#1
{ \hat{b}^{\dagger}_{#2}    } { \hat{b}^{\phantom{\dagger}}_{#2} } } }

\NewDocumentCommand\NP{}{\ensuremath{ N_{\text{p}} }}
\NewDocumentCommand\NAP{}{\ensuremath{ N_{\text{ap}} }}

\DeclareMathOperator{\sign}{sign}

\NewDocumentCommand\albos{ s o }{
    \ensuremath{%
    \Operator{\alpha}^\IfBooleanTF#1{ {\dagger}} { {\vphantom{\dagger} } } \IfNoValueTF{#2}{}{_{#2}}
    }
}

\NewDocumentCommand\bebos{ s o }{
    \ensuremath{%
    \Operator{\beta}^\IfBooleanTF#1{ {\dagger}} { {\vphantom{\dagger} } } \IfNoValueTF{#2}{}{_{#2}}
    }
}

\NewDocumentCommand\Cgeneral{s}{
    \ensuremath{%
    \IfBooleanTF#1{ \Cnosym{} }{\Csym{}}_i
    }
}

\NewDocumentCommand\Cnull{s}{
    \ensuremath{%
    \IfBooleanTF#1{ \Cnosym{} }{\Csym{}}_0
    }
}

\NewDocumentCommand\Cata{m}{
    \ensuremath{%
    {\Gamma}^{\albos*[]\albos[]}_{#1}
    }
}

\NewDocumentCommand\Cbtb{m}{
    \ensuremath{%
    \Gamma^{\bebos*[]\bebos[]}_{#1}
    }
}

\NewDocumentCommand\Catbt{m}{
    \ensuremath{%
    \Gamma^{\albos*[]\bebos*[]}_{#1}
    }
}

\NewDocumentCommand\Cab{m}{
    \ensuremath{%
    \Gamma^{\albos[]\bebos[]}_{#1}
    }
}

\NewDocumentCommand\Catataa{m}{
    \ensuremath{%
    \mathcal{V}^{\albos*[]\albos*[]\albos[]\albos[]}_{#1}
    }
}

\NewDocumentCommand\Catatabt{m}{
    \ensuremath{%
    \mathcal{V}^{\albos*[]\albos*[]\albos[]\bebos*[]}_{#1}
    }
}

\NewDocumentCommand\Catatbtbt{m}{
    \ensuremath{%
    \mathcal{V}^{\albos*[]\albos*[]\bebos*[]\bebos*[]}_{#1}
    }
}

\NewDocumentCommand\Cataab{m}{
    \ensuremath{%
    \mathcal{V}^{\albos*[]\albos[]\albos[]\bebos[]}_{#1}
    }
}

\NewDocumentCommand\Catabtb{m}{
    \ensuremath{%
    \mathcal{V}^{\albos*[]\albos[]\bebos*[]\bebos[]}_{#1}
    }
}

\NewDocumentCommand\Catbtbtb{m}{
    \ensuremath{%
    \mathcal{V}^{\albos*[]\bebos*[]\bebos*[]\bebos[]}_{#1}
    }
}

\NewDocumentCommand\Caabb{m}{
    \ensuremath{%
    \mathcal{V}^{\albos[]\albos[]\bebos[]\bebos[]}_{#1}
    }
}

\NewDocumentCommand\Cabtbb{m}{
    \ensuremath{%
    \mathcal{V}^{\albos[]\bebos*[]\bebos[]\bebos[]}_{#1}
    }
}

\NewDocumentCommand\Cbtbtbb{m}{
    \ensuremath{%
    \mathcal{V}^{\bebos*[]\bebos*[]\bebos[]\bebos[]}_{#1}
    }
}

\definecolor{med_blue}{HTML}{00A3E0}
\definecolor{med_darkblue}{HTML}{0061A0}
\definecolor{light_purple}{HTML}{9954d1}
\definecolor{tu_orange}{HTML}{f28400}
\definecolor{tdo_green}{HTML}{83B818}
\definecolor{tdo_darkgreen}{HTML}{839A00}
\definecolor{our_red}{RGB}{199,92,117}

\theoremstyle{thmstyleone}%
\theoremstyle{thmstyletwo}%

\theoremstyle{thmstylethree}%

\begin{document}

\title[Quantum effects in the magnon spectrum of 2D altermagnets via continuous similarity transformations]{Quantum effects in the magnon spectrum of 2D altermagnets via continuous similarity transformations}


\author[1]{\fnm{Raymond} \sur{Wiedmann}}\email{r.wiedmann@fkf.mpg.de}

\author[2]{\fnm{Dag-Björn} \sur{Hering}}\email{dag.hering@tu-dortmund.de}

\author[2]{\fnm{Vanessa} \sur{Sulaiman}}\email{vanessa.sulaiman@tu-dortmund.de}

\author[3]{\fnm{Matthias R.} \sur{Walther}}\email{matthias.walther@fau.de}

\author[3]{\fnm{Kai P.} \sur{Schmidt}}\email{kai.phillip.schmidt@fau.de}

\author[2]{\fnm{Götz S.} \sur{Uhrig}}\email{goetz.uhrig@tu-dortmund.de}

\affil[1]{\orgname{Max-Planck-Institut für Festkörperforschung}, \orgaddress{\street{Heisenbergstra\ss{}e 1}, \postcode{70569}, \city{Stuttgart},  \country{Germany}}}

\affil[2]{\orgdiv{Condensed Matter Theory, Department of Physics}, \orgname{TU Dortmund University}, \orgaddress{\street{Otto-Hahn-Stra\ss{}e 4}, \postcode{44227}, \city{Dortmund}, \country{Germany}}}

\affil[3]{\orgdiv{Department of Physics}, \orgname{Friedrich-Alexander-Universität Erlangen-Nürnberg (FAU)}, \orgaddress{\street{Staudtstra\ss{}e 7}, \postcode{91058}, \city{Erlangen},  \country{Germany}}}


\abstract{We investigate quantum effects on magnon excitations in a minimal spin-1/2 Heisenberg model for 2D altermagnets on the square lattice. 
A continuous similarity transformation is applied in momentum space to derive an effective Hamiltonian that conserves the number of magnon excitations.
This allows us to quantitatively calculate the one-magnon dispersion, the effects of magnon-magnon interactions, and the dynamic structure factor in a certain range of parameters.
In particular, we focus on the altermagnetic spin splitting of the magnon bands and the size of the roton minimum.
We further map out divergencies of the continuous similarity transformation for different types of generators, which signal either the breakdown of the N\'eel-ordered phase or the presence of significant magnon decay.}


\maketitle


\section{Introduction}\label{s::intro}

Altermagnets are a novel class of collinearly ordered magnetic materials 
which combine features of ferromagnetism and antiferromagnetism~\cite{Smejkal2022Beyond, Smejkal2022Emerging,Mazin2022Editorial}. 
These materials exhibit symmetries acting on spin and lattice degrees of freedom independently~\cite{Yi2024_SSG}, which lead to distinctive electronic and magnetic properties, in particular a large non-relativistic spin splitting in the band structure. Unlike conventional magnetic systems, altermagnets can support spin-polarized currents while maintaining zero net magnetization and without the need for an external magnetic field, which makes them promising candidates for future spintronic applications~\cite{DalDin2024AntiferromagneticSA}. 
Recent advancement in the field has taken place in theoretical investigations~\cite{McClarty2024Landau,Heinsdorf2025,parshukov2024topologicalresponsesgappedweyl}, material predictions~\cite{Bai2024Review,Sodequist2024materialscreening,Xiao2024SSGClassification},  and experimental confirmations of  altermagnetic behavior~\cite{krempasky_jungwirth_nature_24,fedchenko_sci_advances_24,nag2024gdalsiantiferromagnetictopologicalweyl}.

The unique properties of altermagnets also affect the magnetic excitations in these long-range ordered systems, i.e., the magnons, which, similar to antiferromagnets, can carry two distinct spin values $S^z=\pm 1$, sometimes
also called chiralities. 
In contrast to antiferromagnets, the magnons of opposite spin are energetically split in altermagnets 
at general points in the Brillouin zone~\cite{Smejkal2023ChiralMagnons}, making them an intriguing platform for the design of magnonic devices. This spin splitting has been predicted~\cite{mcclarty2024observingaltermagnetismusingpolarized,hoyer2025altermagneticsplittingmagnonshematite} and observed in \gls{ins} experiments~\cite{Liu2024INSMagnonsMnTe}.
For the investigation of magnons in magnetic systems, \gls{lswt} is usually used, neglecting the effects of magnon-magnon interactions in the system completely. 
This is justified in systems with large spin $S$ and large dimension, where the effects of quantum fluctuations are negligible. 
In small-$S$ systems, with the limiting case of spin-$1/2$, these effects have to be taken into account.

The effects of quantum fluctuations and magnon-magnon interactions have, most notably, been studied extensively in the spin-$1/2$ square lattice Heisenberg model
\cite{Powalski15,Powalski18,manousakis_SpintextonehalfHeisenbergAntiferromagnet_1991,
hamer_ThirdorderSpinwaveTheory_1992, singh_SpinwaveExcitationSpectra_1995, sandvik_HighEnergyMagnonDispersion_2001,zheng_SeriesStudiesSpin$frac12$_2005}.
In the case of ferromagnetic interactions, it is well known that the magnons are stable, since the ferromagnetic ground state and the single-magnon excitations are exact eigenstates. 
For general antiferromagnetic Heisenberg models, however, quantum fluctuations play an important role and lead to single-magnon band renormalization and possibly to magnon decay.
A fundamental property in the magnon dispersion, which can only be described with interacting theories,
is the emergence of the roton minimum~\cite{sandvik_HighEnergyMagnonDispersion_2001,zheng_SeriesStudiesSpin$frac12$_2005,Powalski15,Powalski18,verre18b}.
This feature appears in the high-energy part of the magnon spectrum.

The differences in the behavior of magnons in collinear ferro- and antiferromagnets due to quantum fluctuations raise the question of which role magnon-magnon interactions play in altermagnetic systems. 
This is crucial in order to understand whether and how magnons can serve as stable quasi-particles used, for instance, for information storage and transmission in magnonic devices, or to see if interesting novel quantum effects emerge. 
Very recently, these questions have been studied using various methods such as time-dependent matrix product states, tensor networks, density matrix renormalization group, and nonlinear spin-wave theory~\cite{GarciaGaitan2025,eto2025spontaneousmagnondecaysnonrelativistic,costa2024giantspatialanisotropymagnon,Cichutek2025SpontDecay,cichutek2025quantumfluctuationstwodimensionalaltermagnets}. 
It is found that spontaneous magnon decay occurs, which is directionally anisotropic. This is one indication that non-classical effects are important. 
Furthermore, intriguing quantum effects such as the emergence of the roton minimum, as in conventional antiferromagnets, have been predicted~\cite{liu2024}.

A natural step beyond \gls{lswt} results consists in including the magnon-magnon interaction at least on a static mean-field level by resorting to \gls{sct}. 
However, this only leads to some static renormalization of the magnon dispersion, as can be seen in the case of the antiferromagnetic square lattice Heisenberg model~\cite{Powalski15}. 
Nevertheless, the resulting description can serve as a starting point for the method of \gls{cst}.
This method has already proven fruitful for the spin-$1/2$ square lattice XXZ model (including the Heisenberg point) and the frustrated $J_1$-$J_2$ model~\cite{Powalski15,Powalski18,Walther23,Caci24,Hering24}.
The \gls{cst} allows to capture the roton minimum quantitatively and provides spectral densities that include the important spectral shift towards lower energies induced by the attractive magnon-magnon interaction, so that experimental data was understood quantitatively. 
Yet, the critical behavior at quantum phase transitions eluded a quantitative determination.

In the present article, we use the \gls{cst} to study the spin-$1/2$ Heisenberg model with altermagnetic next-nearest neighbor couplings.
The focus of our investigation is the effect of magnon-magnon interaction on the magnon band structure with an emphasis on the effect of the coupling inducing the altermagnetic features.
To this end, we tune this coupling from the parameter regime that stabilizes the \neel{} order to the one that destabilizes it.

We determine the stability of the \neel{} phase and
estimate the range in which we can assume the magnons to be sufficiently stable.
There are analytical arguments~\cite{Cichutek2025SpontDecay,eto2025spontaneousmagnondecaysnonrelativistic} that the upper branch of altermagnetic magnons is always unstable in the sense that it can decay into three magnons of the same or lower energy. Depending on the details, this can also occur for the magnons of the lower branch.
Thus, in a rigorous sense, there are no stable magnons in altermagnets. 
Hence, a \gls{cst} that separates the one-magnon states from the three-magnon sector is likely to fail.
A less ambitious \gls{cst} that only aims at separating the ground state from other magnon sectors, i.e., to eliminate vacuum fluctuation, is not called into question.
The latter scenario is realized with the so-called $0n$ generator, the former by the \gls{qpc} generator.
Thus, inspecting the convergence of the flow induced by different generators informs us about the stability of the vacuum ($0n$ flow converges) and about the stability of the magnons (\gls{qpc} flow converges).

In practice, the issue of stability becomes a quantitative one in energy resolution. 
Is it possible to distinguish a $\delta$-distribution in a spectral function from a narrow Lorentzian?
It must be noted that the kinematically available phase space for magnon decay is fairly small, in particular for small altermagnetic spin splitting.
In this case, the decay rates will be small according to Fermi's Golden rule, and the magnons live fairly long. 
Therefore, they appear to be robust to studies in which the energy resolution is larger, i.e., worse, than the line width of the magnons. 
Our numerical calculations have two limiting factors for the energy resolution: 
(i) Calculations are done for a finite cluster of linear length $L$.
This implies a limit of the energy resolution which can be estimated by $v_\text{spin}/L$ where $v_\text{spin}$ is the spin wave velocity. 
(ii) The continuous basis transformation is performed up to a cutoff of $\ell_\text{max}$, where $\ell$ parametrizes the running couplings and has the unit of inverse energy, see App.\ \ref{a:connection-between-rod-and-energy-resolution}. 
Thus, $1/\ell_\text{max}$ is the other limiting quantity of the energy resolution, so that one has in total a resolution limit of 
\begin{equation}
    \Delta \omega \approx \max(1/\ell_\text{max}, v_\text{spin}/L)\, ,
\end{equation}
where we set $\hbar$ to one henceforth. 
Thus, we will deal with approximately stable, long-lived magnons if their line width is below $\Delta \omega$, i.e., their life time above $1/\Delta \omega$.
This must be kept in mind in the following.

In the regime in which magnons are approximately stable, we study the behavior of the altermagnetic spin splitting of the magnon branches and the roton minimum resulting from magnon-magnon interactions. 
Finally, we calculate the dynamic structure factor, which is directly relevant for \gls{ins} experiments. 

The article is structured as follows: 
First, we introduce the model in \cref{s::model} and subsequently the method in some detail in \cref{s::method}. 
In \cref{s::results} we present and discuss our results on the stability of the \neel{} phase, the quasi-stability of the magnons, and quantitative effects of magnon-magnon interactions on the one-magnon spectrum. 
Finally, we conclude in \cref{s::conclusion} assessing the general influence of the magnon-magnon interactions.


\section{Model}
\label{s::model}
\begin{figure}
    \centering
    \includegraphics[width=\textwidth]{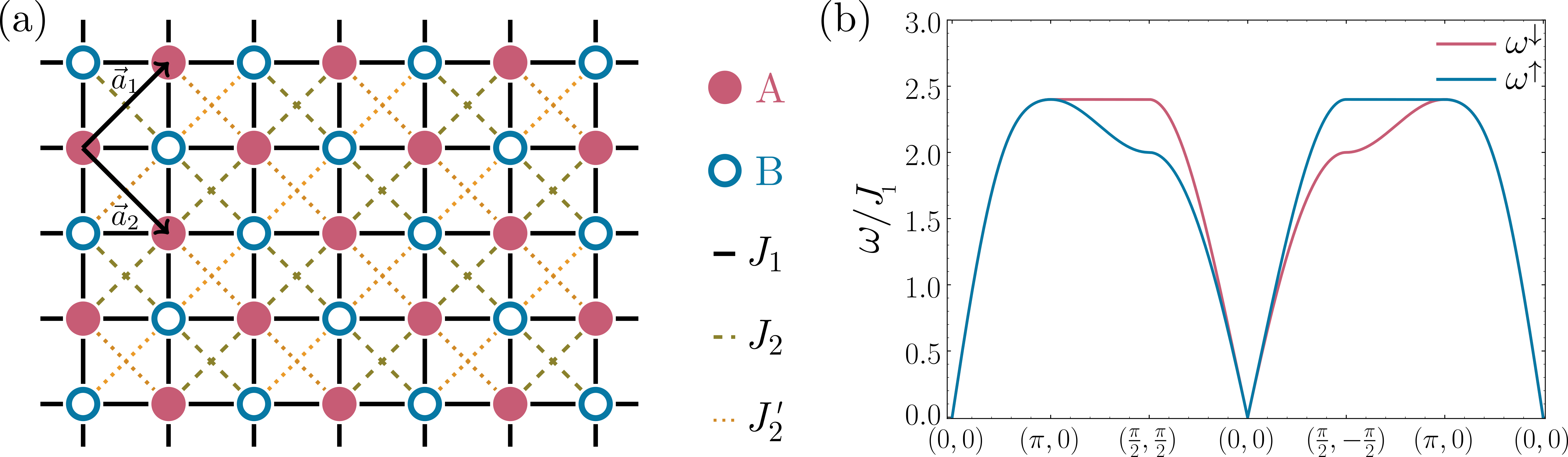}
    \caption{
    (a) Sketch of the spin model \cref{eq::altermagnet} on the square lattice split into the two sublattices A and B, lattice vectors $\Vector{a}_1$ and $\Vector{a}_2$ of the magnetic unit cell, \gls{nn} antiferromagnetic coupling $J_1$, and the \gls{nnn} couplings $J_2^{\vphantom{\prime}}$ and $J_2^\prime$. 
    (b) Magnon bands $\omega^\downarrow$ and $\omega^\uparrow$ obtained from \gls{lswt} in the \neel{} phase for $J_2^{\vphantom{\prime}} = -0.2J_1$, and $J_2^\prime=0$. The lattice constant is set to unity.
    }
    \label{model::altermagnet_lattice}
\end{figure}

We study a minimal spin model for a 2D altermagnet on the square lattice given by~\cite{cui2023,Brekke2023}
\begin{align}\label{eq::altermagnet}
        \mathcal{H} & =  J_1 \sum_{\langle i,j \rangle} S^{(A)}_i \cdot S^{(B)}_j \nonumber\\
                    & + \frac{J_2^{\vphantom{\prime}}}{2} \sum_{ i, \delta \in \pm\Vector{a}_1 } S^{(A)}_i\cdot S^{(A)}_{i+\delta}
                    + \frac{J^\prime_2}{2} \sum_{ i, \delta \in \pm\Vector{a}_2 } S^{(A)}_i\cdot S^{(A)}_{i+\delta} \\
                    & + \frac{J^\prime_2}{2} \sum_{ j, \delta \in \pm\Vector{a}_1 } S^{(B)}_j\cdot S^{(B)}_{j+\delta}
                    + \frac{J_2^{\vphantom{\prime}}}{2} \sum_{ j, \delta \in \pm\Vector{a}_2 } S^{(B)}_j\cdot S^{(B)}_{j+\delta}\ ,\nonumber
\end{align}
where the symbols $S$ stand for the spin vector operators at the sites indicated by the subscripts on one of the two sublattices indicated by the superscripts. Further,  $J_1 > 0$, $J_2^{\vphantom{\prime}}$ and $J_2^\prime$ denote the \gls{nn} and \gls{nnn} couplings of Heisenberg type, respectively, as illustrated in \cref{model::altermagnet_lattice}.
The \gls{nnn} interactions are expressed by sums over the lattice vectors $\Vector{a}_1$ and $\Vector{a}_2$ of the magnetic unit cell. 
The lattice vectors are shown in \cref{model::altermagnet_lattice}(a).
They are identical to the ones of the conventional \gls{nn} antiferromagnetic Heisenberg model on the square lattice. 
Furthermore, the sums either run over the $A$ sublattice for $i$ or over the $B$ sublattice for $j$. 
The prefactors $1/2$ compensate the double counting of the \gls{nnn} couplings.

The case of ${J_2^{\vphantom{\prime}} = J_2^\prime = 0}$ with ${J_1 >0}$ is the usual antiferromagnetic Heisenberg model on a square lattice. 
In this limit, the two magnon bands resulting from the two magnetic sublattices are degenerate in the entire Brillouin zone due to the combination of time reversal and inversion symmetry. 
Introducing a finite isotropic \gls{nnn} coupling ${J_2^{\vphantom{\prime}} = J_2^\prime}$ does not break this combined symmetry so that the bands remain degenerate. 
Depending on the sign, the \gls{nnn} coupling either stabilizes the \neel{} order (ferromagnetic ${J_2^{\vphantom{\prime}}<0}$) or induces frustration (antiferromagnetic ${J_2^{\vphantom{\prime}}>0}$) destabilizing this order. 
The latter corresponds to the $J_1$-$J_2$ Heisenberg model, which has been studied intensely due to its intriguing quantum phase diagram, possibly hosting a valence bond solid and/or spin liquid phases~\cite{Sushkov2001_J1J2,Jiang2012_J1J2,Doretto2014_J1J2,Liu2022_J1J2phasediag}.

By making the \gls{nnn} couplings different ${J_2^{\vphantom{\prime}} \neq J_2^\prime}$, 
the inversion symmetry connecting the two sublattices is broken and reduced to mirror symmetries in $x$ and $y$ direction. 
This leads to split magnon bands at general points in the Brillouin zone and degeneracies at the mirror lines, which is the characterizing feature of altermagnetism. 
The difference in the \gls{nnn} coupling $J_2\neq J_2'$ can be caused by the presence of non-magnetic atoms in the unit cell of the underlying material, being different in the different plaquettes. 
This leads to a reduced symmetry and eventually influences the electronic and magnonic spectrum.
The generic magnon spectrum obtained from \gls{lswt} is shown in \cref{model::altermagnet_lattice}(b), displaying the characteristic band splitting and the gapless magnon modes, where the lattice constant is set to unity here and henceforth. 
For all the following calculations and discussions, we use $J_1=1$ as the energy unit, and we restrict ourselves to $J_2^\prime = 0$ for the simplicity of a one-dimensional space of tuning parameters. 
We tune the altermagnetic spin splitting by varying $J_2^{\vphantom{\prime}}$ where $J_2^{\vphantom{\prime}} = 0$ corresponds to the non-altermagnetic antiferromagnetic Heisenberg case. 
This way to proceed has the advantage that we can easily compare effects in the single-magnon spectra to the well-studied antiferromagnetic Heisenberg case.

\section{Method}
\label{s::method}

\subsection{\Glsentrylong{cst} and its implementation}
\label{ss:cst}

We begin with the classical \neel{} state, i.e., two sublattices with alternating spin up and down.
Next, we express the spin operators by bosons in the Dyson-Maleev representation~\cite{Dyson_1956,malee58b}.
The resulting Hamiltonian is not manifestly Hermitian. 
Subsequently, we apply a self-consistent mean-field decoupling ~\cite{Takahashi_1989, Hida_1990} and a Fourier transformation. 
Then, we apply a Bogoliubov transformation to diagonalize the bilinear part of this Hamiltonian and also express all interactions in terms of the Bogoliubov bosons, leading to the Hamiltonian given in App.~\ref{a:self-consisten-mft-altermagnetic-interaction}.
The Bogoliubov bosons display a gapless dispersion so that they comply with the Goldstone theorem, applicable here due to the spontaneously broken continuous spin symmetry.
We stress that this form of the Hamiltonian is still rigorously exact and can be written down in the thermodynamic limit; no approximation has been made so far.
Finally, the Hamiltonian is subjected to the previously used approach of \gls{cst}~\cite{Powalski15,Powalski18,Walther23,Hering24}.

Due to the non-Hermiticity of the Dyson-Maleev Hamiltonian, we have to employ a \gls{cst} instead of the more often used \glspl{cut}~\cite{Wegner_1994,Mielke_1998,Knetter_2000,Knetter_2003}.
Nonetheless, the main idea is the same: A given initial Hamiltonian $\mathcal{H}_0$ is transformed in a continuous way into a more (block-)diagonal form $\mathcal{H}_\text{eff}$.
Given a chosen generator $\eta(\ell)$, the continuous transformation is given by the solution of the flow equation $\partial_{\ell} \mathcal{H}(\ell) = \left[ \eta(\ell), \mathcal{H}(\ell) \right]$.
For $\ell\to\infty$ we arrive at the block-diagonal Hamiltonian $\mathcal{H}(\ell=\infty) = \mathcal{H}_\text{eff}$, which is simpler than the initial Hamiltonian.

Computing the flow equation requires a suitably chosen generator. Also, a truncation scheme is needed because the commutator $\left[ \eta(\ell), \mathcal{H}(\ell) \right]$  generally yields an infinite series of arbitrarily complicated terms.
In this work, we use the $0n$ generator and the \gls{qpc} generator.
The $0n$ generator~\cite{Fischer_2010} disentangles the ground state from all sectors with finite number of quasi-particles, here magnons.
The flow induced by this generator is generally robust, as energetic overlaps among higher magnon sectors are irrelevant since those sectors are not diagonalized. The quantum phase transition out of the ordered phases can be investigated by inspecting the stability of the flow generated by this generator.

In contrast, the \gls{qpc} generator disentangles all quasi-particle sectors so that each block of the block-diagonal effective Hamiltonian $\mathcal{H}_\text{eff}$ contains only elements with a fixed number of quasi-particles, here magnons.
However, the flow induced by this generator only converges if there are no energetic overlaps between different quasi-particle sectors.
For instance, if an eigen energy in the one-quasi-particle (1QP) sector overlaps with the ground-state energy (0QP) sector, i.e., 
it falls below the ground state energy, this signals the closure of the single-particle gap, indicating a second-order quantum phase transition. 
Similarly, overlaps involving sectors with higher quasi-particle numbers can also occur above a stable phase, implying 
a divergent \gls{qpc} flow. Such energetic overlaps can result from binding effects or overlapping continua of scattering states.
Energetic overlaps and the resulting flow divergence are phenomena we indeed encounter
in altermagnets and will examine in detail in \cref{ss::convergence}.

The truncation scheme we use is based on the scaling dimension~\cite{Powalski15, Powalski18, Walther23, Caci24, Hering24} to include the most relevant magnon-magnon interactions in a systematic way
because operators with a large scaling dimension $\dsc$ are less relevant than operators with low scaling dimension, especially for gapless phases. 
The \gls{cst} with the \gls{qpc} generator and a truncation of operators with $\dsc > 2$ were used to quantitatively determine the one-magnon dispersion and the dynamical structure factor of the magnetically ordered \neel{} phase in the antiferromagnetic spin-$1/2$ Heisenberg model on the square lattice~\cite{Powalski15,Powalski18} as well as the single- and two-particle properties in the easy-axis spin-$1/2$ XXZ model on the square lattice~\cite{Walther23,Caci24} quantitatively.
Furthermore, the quantum critical point of the phase transition between magnetically ordered phases and non-magnetic phases has been determined in the Heisenberg bilayer and $J_1${-}$J_2$ model with good accuracy with the $0n$ generator.
The determination of critical exponents proved to be challenging in these models~\cite{Hering24}.

In order to solve the flow equation in momentum space numerically, we discretize the \gls{bz} with $L^2$ equidistant points, where $L$ denotes the linear system size. 
The discretization can be done in two distinct ways, where the crucial difference lies in the treatment of the $\Gamma$ point 
$\kvec{}=\left(0,0\right)$. 
For the \gls{pbc}, the $\left(0,0\right)$ point is included and we label
these discretizations by \acrshort{pbc}.
In contrast, \gls{apbc} avoid the $\left(0,0\right)$ point automatically.
The $\left(0,0\right)$ point is numerically challenging since at this point the mean-field solution displays an integrable divergence.
Therefore, for \gls{pbc}, the coefficients pertaining to the $\left(0,0\right)$-point are set to zero, 
effectively switching them off during the flow.
In the following, the mesh points for \gls{pbc} are denoted by $\NP{}$ and the ones for \gls{apbc} by $\NAP{}$.

The rigorous limit $\ell\to\infty$ is not reachable numerically.
Hence, the flow is stopped in the calculations when the \gls{rod}, i.e., 
the square root of the sum over all squared moduli of the coefficients occurring in the corresponding generator 
$\eta$~\cite{Fischer_2010}, drops to values below $10^{-6}J_1$. 
This leads to a finite energy resolution depending on $1/\ell_{\text{max}}$ 
which is discussed below and in App.~\ref{a:connection-between-rod-and-energy-resolution}.
We choose the \gls{rod} threshold independently of the system size.
Due to the quickly increasing number of coefficients $\propto L^6$ for larger systems, the requirement that the \gls{rod}
falls below the same numerical values represents a stricter threshold, the larger the system is.
Thus, the coefficients are reduced even further as the number of entries grows, yielding a higher numerical accuracy.

\subsection{Calculation of spectral densities}
\label {ss:sd}

To compute observables in the \gls{cst} framework, their operators have to be transformed in the same manner as the Hamiltonian~\cite{Powalski18}
in order to represent them in the same basis.
If the Hamiltonian can be transformed using the \gls{qpc} generator, this has the significant benefit of the 
separation of the sectors with different magnon numbers. 
The calculation of observables and of their dynamics can be done in each subsector separately.

Generically, magnon bands are measured using \acrfull{ins}, which has recently been employed to demonstrate the spin-split magnon bands in the altermagnet MnTe~\cite{Liu2024INSMagnonsMnTe}. 
Since \gls{ins} probes the bulk properties of a material, measuring the magnon bands of a 2D system is only possible in a 3D layered structure with negligible interlayer coupling, e.g., quasi-2D van der Waals magnetic materials~\cite{Zhu2021_INS2DvdW}. 

In \gls{ins} experiments, a frequency and momentum dependent counting rate $I(\omega,\Vector{Q})$ is measured.
For sufficiently low temperatures, it is proportional to the dynamic structure factor (DSF) at $T=0$
\begin{equation}
    S^{\alpha\alpha}(\omega, \Vector{Q}) = -\frac{1}{\pi}\Im \bra{0}S_\text{eff}^\alpha(-\Vector{Q}) \frac{1}{\omega-(\mathcal{H}_\text{eff}-\mathcal{E}_0)} S_\text{eff}^\alpha(\Vector{Q})\ket{0}~.
\end{equation}
Here, $\ket{0}$ is the ground state of $\mathcal{H}_\text{eff}$ and $\mathcal{E}_0$ its ground-state energy.
The $S_\text{eff}^\alpha(\Vector{Q})$ are the transformed components of the spin operator with $\alpha\in\{ x,y,z\}$.
Using polarized neutrons, longitudinal ($\alpha = z$) and transversal ($\alpha = x,y$) contributions to the DSF
can be distinguished. The total $S^z$ component takes the values $\pm1$ for the transversal modes and $0$ for the longitudinal one~\cite{mcclarty2024observingaltermagnetismusingpolarized}. 
The transversal contribution consists of two parts
\begin{subequations}
\begin{align}
    S^{xx+yy}(\omega,\Vector{Q}) 
    &= S^{xx}(\omega,\Vector{Q}) + S^{yy}(\omega,\Vector{Q}) \\
    &= S^{+-}(\omega,\Vector{Q}) + S^{-+}(\omega,\Vector{Q}) \\
    \begin{split} 
    &=-\frac{1}{\pi}\Im \bra{0}S^+(-\Vector{Q}) \frac{1}{\omega-(\mathcal{H}_\text{eff}-\mathcal{E}_0)} S^-(\Vector{Q})\ket{0} \\
    &\phantom{=\,}-\frac{1}{\pi}\Im \bra{0}S^-(-\Vector{Q}) \frac{1}{\omega-(\mathcal{H}_\text{eff}-\mathcal{E}_0)} S^+(\Vector{Q})\ket{0} 
    \end{split}
\end{align}    
\end{subequations}
as either the $\omega^\downarrow$ mode ($S^{+-}$) or the $\omega^\uparrow$ mode ($S^{-+}$) is excited.
The transversal component generally couples to the blocks with odd magnon number ($n=1, 3, \ldots$) 
while the longitudinal component couples to even magnon number sectors ($n=2, 4, \ldots$).
We do not consider sectors with more magnons in accordance with the truncation scheme used for the Hamiltonian.

Numerically, the resolvent
\begin{equation}
    R(\omega,\Vector{Q}) = \bra{0}S_\text{eff}^\alpha(-\Vector{Q}) \frac{1}{\omega-(\mathcal{H}_\text{eff}-\mathcal{E}_0)} S_\text{eff}^\alpha(\Vector{Q})\ket{0}
\end{equation}
can be calculated efficiently via a Lanczos algorithm, which provides a continued fraction representation of the resolvent.
Due to the non-Hermiticity of the Hamiltonian, we have to use a non-symmetric, 
biorthogonal Lanczos algorithm~\cite{doi:10.1137/1.9780898719581.ch7,Powalski18}.
With the start vectors ${\bra{v_L} = \bra{0}S_\text{eff}^\alpha(-\Vector{Q})}$ and ${\ket{v_R} = S_\text{eff}^\alpha(\Vector{Q})\ket{0}}$ this biorthogonal Lanczos algorithm provides the continued fraction
\begin{align}
    R(\omega)
    = \frac{\Braket{v_L|v_R}}{\omega-\alpha_0-\frac{\beta_1\gamma_1}{\omega-\alpha_1-\frac{\beta_2\gamma_2}{\dots}}}
    = \frac{\beta_0\gamma_0}{\omega-\alpha_0-\frac{\beta_1\gamma_1}{\omega-\alpha_1-\frac{\beta_2\gamma_2}{\dots}}}
\end{align}
with $\alpha_i$ being the diagonal elements of the resulting tridiagonal matrix and $\beta_i$ and $\gamma_i$ the off-diagonal elements.

\section{Results}
\label{s::results}

\subsection{Stability of the phase and magnon decay}
\label{ss::convergence}
The two generator schemes for the \gls{cst} -- $0n$ and \gls{qpc} -- can be used to understand the stability of the \neel{} phase and the relevance of magnon decay in the considered altermagnetic system.
As explained in \cref{s::method}, the \gls{qpc} generator aims to bring the initial Hamiltonian into a block-diagonal form, where the number of magnons in each block is constant, while a \gls{cst} based on the $0n$ generator just disentangles the ground-state block from all other single- and multiple-particle blocks.

First, we employ the $0n$ generator in order to map out the boundaries of the \neel{} phase depending on the coupling parameter $J_2$. 
For this, we use that the flow based on the $0n$ generator only converges if the magnon vacuum remains the ground state. 
This means that the ground state is adiabatically connected to the \neel{} phase without a phase transition.
Consequently, we can use the convergence of the flow to find the boundaries of the \neel{} phase in dependency on $J_2$.
We find that for $J_2<0$, i.e., ferromagnetic \gls{nnn} coupling, the $0n$ flow always converges. 
This was to be expected since ferromagnetic couplings between spins of the same sublattice stabilize the \neel{} order 
and no phase transition should occur.
Note that this also holds when including finite negative $J_2^\prime$.

The situation is different for the frustrated antiferromagnetic case ($J_2 > 0$), where the \gls{nnn} coupling 
competes with the NN antiferromagnetic coupling.
For $J_2^{\vphantom{\prime}} = J_2^\prime > 0$ the model becomes the well-studied $J_1${-}$J_2$ model for which 
various methods predict a phase transition to a non-magnetic phase in a range $J_{2,c}^{\vphantom{\prime}}/J_1 = J_{2,c}^\prime/J_1 \in [0.35,0.45]$~\cite{richter_Spin12J1J2_2010,morita_QuantumSpinLiquid_2015,gong_PlaquetteOrderedPhase_2014,liu_GaplessQuantumSpin_2022,bishop_FrustratedSpin1_2019}. For this model, the \gls{cst} is known to capture the breakdown of the magnetically ordered phase at 
$J_2^{\vphantom{\prime}} = J_2^\prime \approx 0.372J_1$~\cite{Hering24}.

\begin{figure}
   \centering
   \includegraphics[width=0.8\textwidth]{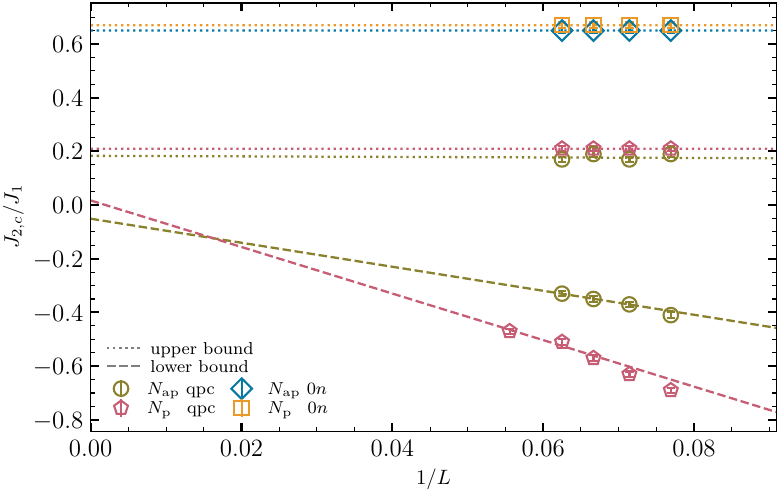}
   \caption{
   Linear extrapolation of the different endpoints of convergence for the analyzed parameter regime with $J_2'=0$.
   The convergence endpoints for both the $0n$ and the \gls{qpc} generator are shown depending on the inverse linear system size $1/L$.
   For $0n$, only an upper set of endpoints is found.
   }     
   \label{fig::Convergence}
\end{figure}

To determine the quantum critical point $J_{2,\text{c}}$ for $J'_2=0$ via the $0n$ generator we use the same technique as previously~\cite{Hering24} for both boundary conditions, \gls{pbc} and \gls{apbc}.
In \cref{fig::Convergence} the critical $J_{2,\text{c}}$ are shown for different linear system sizes $L\in\{13,14,15,16\}$ 
and both boundary conditions.
A linear extrapolation yields  $J_{2,\text{c}}\approx(\num{0.66(1)})J_1$.
Compared to the results of the $J_1$-$J_2$ model,  $J_{2,\text{c}}$ is almost twice as large. 
This can be explained by the absence of $J^\prime_{2}$, which requires that the destabilizing 
magnetic coupling be incremented to compensate for the lack of $J^\prime_{2}$.

Similar to the convergence of the $0n$ generator, the flow of the \gls{qpc} generator only converges if the energy of any mode with $n$ magnons
is always smaller than or equal to the energy of a mode with $m>n$ magnons within the truncation scheme used.
This means that the divergence of the \gls{qpc} flow indicates a level-crossing between modes with a different number of magnons.
Since we are considering quartic terms at most, only the one-magnon energies overlapping with three-magnon energies can be at the origin of such a divergence. 
Hence, a decay channel for single magnons is indicated.
In fact, recent studies demonstrated by kinematic analysis that altermagnets generally allow for spontaneous single-magnon decay~\cite{eto2025spontaneousmagnondecaysnonrelativistic,Cichutek2025SpontDecay}. 
In particular, the upper magnon branch is always instable~\cite{eto2025spontaneousmagnondecaysnonrelativistic}
with an anisotropic decay rate in the Brillouin zone. 
While the analytic arguments concentrate on the long-wavelength limit, considering continuum models, the \gls{cst} is performed 
in the full \gls{bz}, with resolution restricted by the discretization in momentum space. 
Additionally, the maximum running parameter  $\ell_\text{max}$ 
indicates a second restriction of the resolution, see
App.~\ref{a:connection-between-rod-and-energy-resolution}.

Therefore, magnon decay cannot be captured with arbitrary accuracy so that sharp Lorentzians and $\delta$ distributions
cannot be told apart. In this sense, we use the term of approximate stability and map out the parameter region in which the \gls{qpc} generator converges, providing 
an effective model in terms of approximately stable magnons.

To determine the boundaries for which the \gls{qpc} generator converges, we proceed in the same manner as for the $0n$ generator. 
Contrary to the $0n$ generator, a converging flow is no longer ensured for $J_2<0$ and we find both
an upper and a lower bound for the convergence area of the \gls{qpc} generator.
We observe a significant difference in the scaling of the two bounds. 
The lower bound $J^\text{l}_{2,\text{c}}$ shows a greater dependence on the system size compared to the upper bound $J^\text{u}_{2,\text{c}}$.
The extrapolation of the lower boundary yields almost zero with $J^\text{l}_{2,\text{c}}=(\num{0.02(3)}) J_1$ while the 
extrapolated upper boundary stays finite with $J^\text{u}_{2,\text{c}}=(\num{0.20(2)}) J_1$.
Note that the lower point of divergence of the \gls{qpc} generator is far more stable for \gls{apbc} than for \gls{pbc}. 
To obtain a good extrapolation, nevertheless, an additional cluster size with $L=18$ is evaluated for \gls{pbc}. 

Even after extrapolation, our results show that the \gls{qpc} flow converges in an extended parameter region with finite altermagnetic splitting, indicating approximate stability of the magnons.
Therefore, we treat the magnons as stable quasi-particles, which allows us to discuss the interaction effects on the single-magnon spectrum in the following section.
We stress that we find a boundary in the case of $J_2>0$, which is almost independent of the discretization. 
We attribute this unexpected observation to the frustration induced by competing antiferromagnetic couplings.

\subsection{Single-magnon properties}
\label{s::single_magnon}

In this and the following section, we discuss the influence of the magnon-magnon interactions on two important properties of the dispersion of the altermagnetic model in \cref{eq::altermagnet} and the spin-$1/2$ antiferromagnetic Heisenberg model on the square lattice, respectively: the altermagnetic spin-splitting $\deltaSpin$ and the depth of the roton minimum $\Delta R$.

To isolate the effects of the magnon-magnon interaction from effects already captured by \gls{lswt} or \gls{sct}, the considered case of $J_2^\prime = 0$ is well suited.
Then, we choose a high-symmetry path where the shape of the $\omega^\downarrow$ mode is not strongly affected by the \gls{nnn} coupling in \gls{lswt} and \gls{sct} as can be seen in \cref{model::altermagnet_lattice}.
In particular, along the high-symmetry line $(\pi,0)$ to $(\pi/2,\pi/2)$ the mode exhibits the same plateau as in the antiferromagnetic Heisenberg model, but renormalized by the factor $(J_1 - J_2)/J_1$. 
Based on this observation, we can attribute any changes in the dispersion of the $\omega^\downarrow$ mode between $(\pi,0)$ and $(\pi/2,\pi/2)$ to magnon-magnon interactions and compare the effects of (anti)ferromagnetic \gls{nnn} interactions with the usual antiferromagnetic NN exchange.
The dispersion of the $\omega^\uparrow$ mode along the same path is substantially changed, with the largest energy difference to the $\omega^\downarrow$ mode arising at $(\pi/2,\pi/2)$.
The energy difference between both modes at this momentum is denoted as the spin splitting $\deltaSpin$ in the following.

Note that due to the four-fold rotational symmetry connecting the A and B sublattices, the behavior of the two modes swaps in different quadrants of the \gls{bz}. 
As shown in \cref{model::altermagnet_lattice}, the $\omega^\uparrow$ mode 
exhibits the plateau between $(\pi/2,-\pi/2)$ and $(\pi,0)$.
Due to this symmetry, it is sufficient to study only a part of the \gls{bz}.
We focus on the first half of the high-symmetry path where the $\omega^\downarrow$ mode displays the plateau.

As discussed in \Cref{ss::convergence}, a converging \gls{qpc} generator is required to study the single-magnon sector for finite coupling $J_2$.
Although the extrapolated boundaries of the convergence region of the \gls{qpc} generator
in \cref{ss::convergence} suggest a range of $0\lesssim J_2/J_1 \lesssim 0.2$, we 
choose a symmetric range in this section around $J_2=0$ with $\lvert J_2\rvert \leq 0.16 J_1$
to illustrate the effects of various values of the \gls{nnn} coupling.
This range is located between the endpoints of convergence of the finite systems under study.
At these values of the \gls{nnn} coupling, the endpoints only display a limited dependence on the system size $L$.

As in previous studies~\cite{Walther23,Caci24,Hering24} results in the thermodynamic limit $L\to\infty$ 
can be obtained via linear extrapolations of relevant points of the single-magnon dispersion in $1/L$.
In general, the results of \gls{apbc} show less dependency on the system size compared to results for \gls{pbc}, thus we use the results for \gls{apbc}  as the final result here.  
The difference in the thermodynamic limit between the two boundary conditions provides an error estimate for the result.
A notable technical difference compared to previous studies is that missing points in the dispersion due to incompatible discretizations are retrieved by means of regular grid interpolation.

\begin{figure}
    \includegraphics[width=\linewidth]{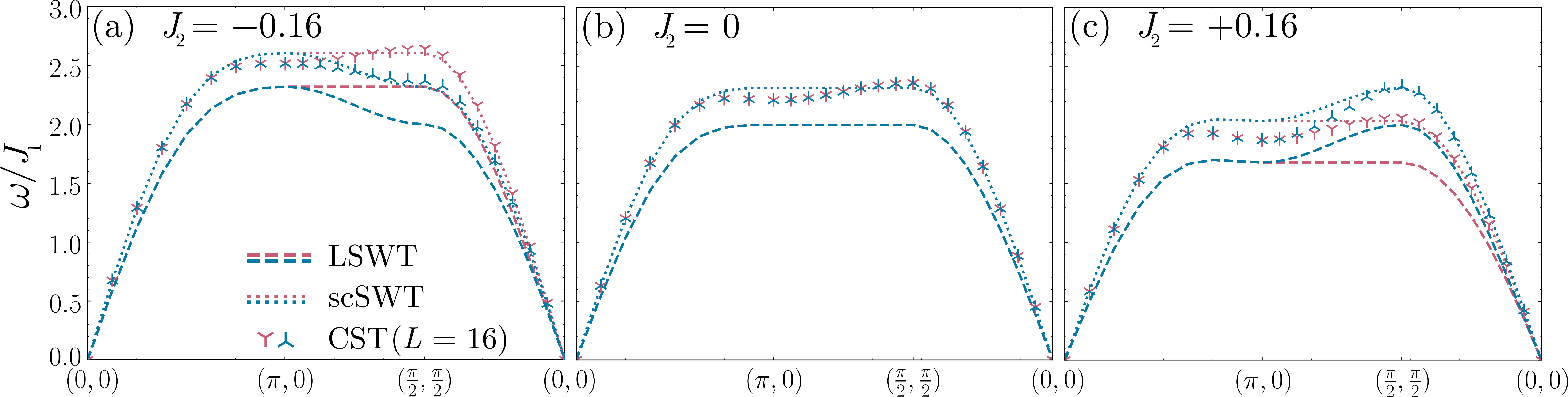}
    \caption{
    Dispersions of the $\omega^\downarrow$ and $\omega^\uparrow$ magnon mode obtained from 
    \gls{lswt} (dashed lines), \gls{sct} (dotted lines) and \gls{cst} ($L=16$; crosses) 
    along a high-symmetry path for (a) $J_2 = -0.16$, (b) $J_2 = 0$, and (c) $J_2 = +0.16$ for $J_2^\prime = 0$.
    }
    \label{fig::raw_disperions}
\end{figure}

\cref{fig::raw_disperions} shows the result for \gls{lswt}, \gls{sct}, and \gls{cst} ($L=16$) for $J_2=0$ and $J_2 = \pm 0.16J_1$.
The characteristic plateau at $2(J_1 - J_2)$ in \gls{lswt} of the $\omega^\downarrow$ magnon is obvious 
for all cases and the renormalization of the dispersion in \gls{sct}. 
The \gls{nnn} interactions cause a positive (negative) spin-splitting for ferro- 
(antiferro-)magnetic $J_2$
\begin{equation}
    \deltaSpin = \omega^\downarrow ( \pi/2,\pi/2 ) - \omega^\uparrow ( \pi/2,\pi/2 )\ ,
\end{equation}
which we analyze in the following section.

Apart from the altermagnetic spin splitting, we find the roton minimum at the $(\pi,0)$ point in the \gls{cst} data, which is not present in \gls{lswt} and \gls{sct} data.
The roton minimum cannot be described without taking quantum effects into account. 
It has been studied in the antiferromagnetic Heisenberg model and is understood to be caused by the 
attractive magnon-magnon interaction shifting spectral weight downwards in energy, thereby
leading to an enhanced level repulsion effect between the single-magnon dispersion and the three-magnon continuum~\cite{Powalski15,Powalski18}. 
In real space, it was understood as a perturbative effect on magnon propagation along diagonals via a three-magnon virtual state~\cite{verre18b}.
The antiferromagnetic Heisenberg model is a limiting case of the altermagnetic model, for $J_2=0$, as shown in \cref{fig::raw_disperions}(b). 
The roton minimum corresponds to the energetic difference between a dip and a peak in energy emerging from the plateau at the $(\pi,0)$ and $(\pi/2,\pi/2)$ points, respectively. 
Therefore, we quantify the roton minimum by this energy difference
\begin{align}
\label{eq:roton}
    \Delta R = \omega^\downarrow ( \pi/2,\pi/2 ) - \omega^\downarrow ( \pi, 0)\
\end{align}
in the $\omega^\downarrow$ mode, which is exactly zero in \gls{lswt} even 
upon including the \gls{nnn} coupling $J_2$. 
In the following, we will discuss the dependence of 
the characteristic roton minimum and altermagnetic spin splitting
in the magnon spectrum on the \gls{nnn} coupling $J_2$.

\subsubsection{Altermagnetic spin splitting}

\begin{figure}[ht!]
    \centering
    \includegraphics[width=0.8\textwidth]{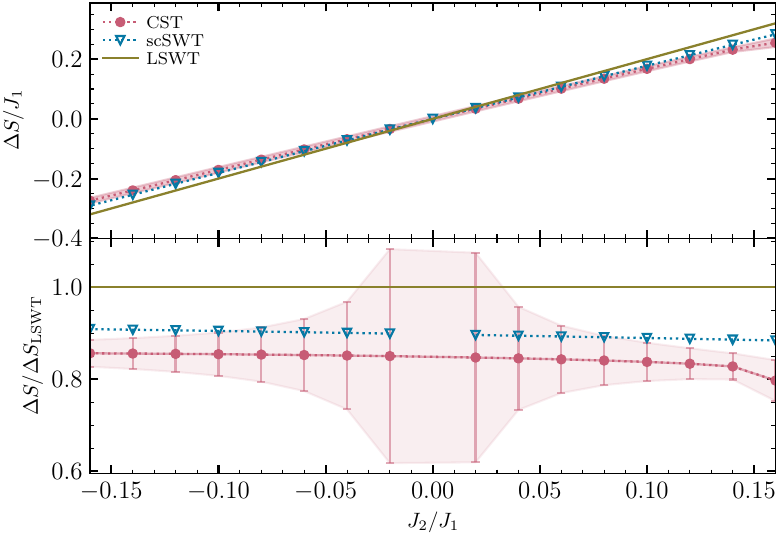}
    \caption{
    The upper panel depicts the altermagnetic spin splitting $\Delta S$ for different $J_2$, 
    where $J_2<0$ ($J_2>0$) corresponds to ferromagnetic (antiferromagnetic) \gls{nnn} coupling.
    The lower panel shows the spin splitting in \gls{sct} and \gls{cst} relative to \gls{lswt}.
    }
    \label{results::magnon_splitting}
\end{figure}

The non-relativistic spin splitting is the main feature in the band structures of altermagnets and holds great promises for applications using spin-polarized electronic and magnonic currents for spintronics and magnonics, respectively.
Therefore, the stability of this splitting is of great interest.
In order to study the effect of magnon-magnon interactions on the altermagnetic spin splitting of the two magnon bands, we compare \gls{lswt} (taking no interaction effects into account), \gls{sct} (including interactions on a static mean-field level), and \gls{cst} (fully including magnon-magnon interactions) for various values of
the \gls{nnn} coupling $J_2$. The results are shown in \cref{results::magnon_splitting}.

We use the symmetric range of $J_2$, in which \gls{cst} converges for $L=16$ as discussed in the previous section. Since we consider the case of $J_2^\prime=0$, the size of the splitting in \gls{lswt} is given by
$\deltaSpin = 2J_2/J_1$, i.e., it is proportional to $J_2$. 
For a quantitative comparison, we normalize all results by this factor, which implies a constant 
relative splitting $\deltaSpin/\deltaSpin{}_{\text{LSWT}}$ in \gls{lswt} by construction, see lower panel of \cref{results::magnon_splitting}. We stress that the large error bars close to $J_2=0$
ensue from the relative normalization.

In \gls{sct}, we observe a reduction of the spin splitting between 
$\SI{9}{\percent}$ and $\SI{12}{\percent}$ depending on $J_2$. 
On the one hand, the deviation from \gls{lswt} is slightly smaller for negative values of the coupling. 
This corresponds to the case of strong ferromagnetic interactions between the next-nearest neighbors 
which stabilizes the magnetic order.
For $J_2>0$, on the other hand, the antiferromagnetic nearest and next-nearest neighbor interactions compete, 
leading to a stronger decrease of the spin splitting in the magnon spectrum.
This indicates that the stabilizing or destabilizing character of the \gls{nnn} coupling 
$J_2$ affects the size of the spin splitting already on a mean-field level to some extent.

The full interaction effects are captured in the \gls{cst} calculation, where we find an additional decrease of the spin splitting in comparison to the \gls{sct} result.
The behavior with respect to the sign and size of $J_2$ is similar. 
For large $J_2$, however, close to the limit of convergence, the size of the splitting drops nonlinearly, which can be ascribed to the effects of competing antiferromagnetic interactions.
We emphasize that the enhanced quantum fluctuations have a stronger influence on the splitting than what is
captured by \gls{sct}.

Summarizing, the decrease of the altermagnetic spin splitting captured by \gls{cst} compared to \gls{lswt} lies 
in the analyzed range within $\SI{14}{\percent}$ to $\SI{20}{\percent}$. Thus, the magnon-magnon interactions at $T=0$ have a quantitative effect on the spectrum. Yet, the qualitative feature of a spin splitting
is a robust signature of altermagnetism in the considered parameter range.

\subsubsection{Depth of roton minimum}

\begin{figure}[ht!]
    \centering
    \includegraphics[width=0.8\textwidth]{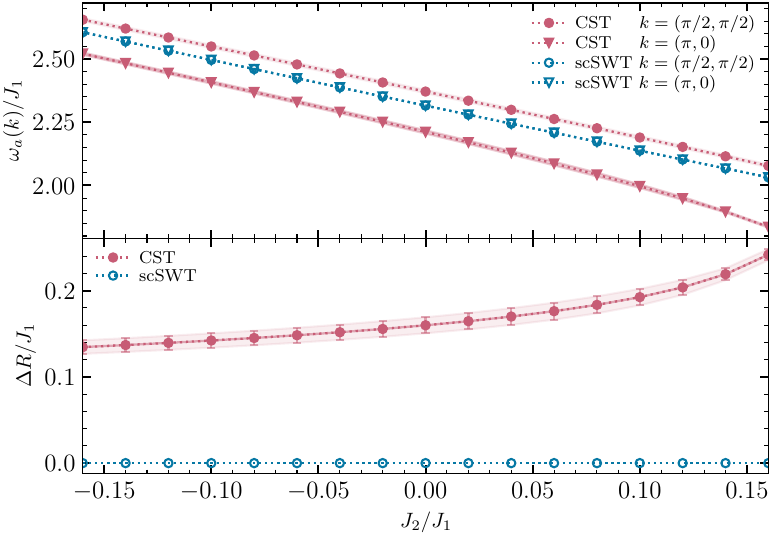}
    \caption{
    The upper panel shows the dispersion $\omega^\downarrow(k)$ for the two point $k=\left(\pi/2,\pi/2\right)$ (circles) and $k=\left(\pi,0\right)$ (triangles).
    The lower panel shows the resulting roton minimum $\Delta R$ for different $J_2$, where $J_2<0$ ($J_2>0$) corresponds to ferromagnetic (antiferromagnetic) next-nearest neighbor coupling.
    Results are shown for both \gls{sct} and \gls{cst}.
    }
    \label{results::roton_depth}
\end{figure}

As described above, the roton minimum is a feature emerging in the high-energy part of the magnon spectrum of an antiferromagnetic Heisenberg model which cannot be captured by \gls{lswt} or \gls{sct}.
It arises from the attractive magnon-magnon interaction leading to an enhanced level repulsion between
the single-magnon dispersion and the three-magnon continuum. Therefore, it represents a fully non-classical effect.

Since the model under study corresponds to the antiferromagnetic Heisenberg model for $J_2 = J_2^\prime = 0$, the question of the fate of the roton minimum under the inclusion of the \gls{nnn} couplings arises naturally.
A recent study~\cite{liu2024} provided first evidence of the roton minimum in the altermagnetic case.
There, however, for the chosen parameters, the \gls{nnn} interaction already causes the energetic dip in \gls{lswt} and the pure quantum effects have not been discussed so far.

In order to isolate the quantum effects leading to the energetic dip in the magnon bands, we
consider the $\omega^\downarrow$ mode between $(\pi, 0)$ and $(\pi/2,\pi/2)$
since its dispersion for $J_2^\prime=0$ is not affected by the \gls{nnn} coupling in \gls{lswt} nor in \gls{sct} except for a shift of the antiferromagnetic Heisenberg plateau, see
\cref{fig::raw_disperions}.
In analogy to the altermagnetic spin splitting, we study the dependence of the roton minimum
as a function of $J_2$.
In \cref{results::roton_depth}, the upper panel depicts the absolute values of the dispersion $\omega^\downarrow(k)$ at the points $(\pi,0)$ and $(\pi/2,\pi/2)$ for \gls{sct} and \gls{cst}, and the lower panel shows the roton minimum $\Delta R$ as defined in Eq.~\eqref{eq:roton} obtained from \gls{sct} and \gls{cst}.

As expected, the \gls{sct} results do not capture the quantum effect of the roton minimum, 
i.e., $\Delta R = 0$ for any $J_2$.
Using \gls{cst}, we obtain a finite roton minimum in the $\omega^\downarrow$ mode, which strongly depends on $J_2$.
For $J_2<0$, one finds a slight decrease of $\Delta R$ upon passing to larger negative values. 
This aligns with the understanding that a ferromagnetic coupling of the next-nearest neighbors stabilizes the 
\neel{} order, thereby suppressing quantum fluctuations.
In the case of $J_2>0$, i.e., frustrating antiferromagnetic coupling, the order is increasingly destabilized upon increasing the frustrating coupling $J_2$. This enhances quantum fluctuations and leads to a strong increase in $\Delta R$. 
We stress that the observed dependence of the roton minimum on the \gls{nnn} coupling is not a classical dispersion effect because $\Delta R$ is zero in the \gls{sct} case for any coupling parameters.
Hence, it is a distinct quantum effect that we find to be impacted non-trivially by the coupling
inducing altermagnetism.
In particular, in the antiferromagnetic Heisenberg model, the roton minimum scales linearly with the coupling strength $J_1$. This coupling also determines the height of the plateau in the \gls{lswt} band structure.

One might expect a similar behavior for the altermagnetic model.
A scaling of the roton depth proportional to the plateau value in the altermagnetic model 
would imply $\Delta R = \Delta R^{\text{HB}} (J_1 - J_2)/J_1$ where
$\Delta R^{\text{HB}}$ is the roton depth of the NN Heisenberg model.
In contrast to this expectation, we find that the behavior is not only nonlinear, 
especially for $J_2>0$, but also of inverse monotonicity: the roton depth increases upon increasing $J_2$.
In particular, inspecting the upper panel in~\cref{results::roton_depth}, 
the mode at $(\pi, 0)$ is more strongly influenced by $J_2$ than the mode at $(\pi/2, \pi/2)$.
Therefore, our findings cannot be explained by such a trivial scaling, and the effects of the \gls{nnn} 
coupling on the quantum fluctuations have to be taken into account.
For spin-$1/2$ systems at very low temperatures, this effect needs to be taken into account 
when studying the magnon spectrum. 
The roton depth can serve as an indicator for the relevance of quantum fluctuations in the system.

\subsection{Spectral densities}

We calculate the magnon spectral densities using \gls{cst}, which allows us to separate the channels of different magnon numbers as explained in \cref{ss:sd}. 
We recall  that this is only possible in the parameter region where different subspaces do not overlap essentially in energy, i.e., 
the area of convergence of the \gls{qpc} generator mapped out above.

We only use \acrlong{pbc} because any sum of momenta is again a valid point 
of the discretization mesh of the \gls{bz}.
For clarity, we point out that this is not the case for \gls{apbc}, where sums of an even number of summands are not in the discretization mesh.
This would be the case for the momenta of the longitudinal contribution ($S^{zz}\left(\omega,\Vector{Q}\right)$) with $\Vector{Q}\in\NP{}$.
Thus, the momenta could not be chosen to be the same as for the transversal contributions
$(S^{+-}\left(\omega,\Vector{Q}\right),S^{-+}\left(\omega,\Vector{Q}\right))$ with $\Vector{Q}\in\NAP{}$. 
To avoid such a mixture of different momenta, we only analyze the results for \gls{pbc}.
Furthermore, we use a fixed linear system size of $L=16$.
To plot the dynamical structure factors, the continued fraction is broadened through an imaginary part of $J_1/10$, i.e., each frequency is shifted by $\omega\to\omega+{\rm i}\,0.1 J_1$ into the complex plane.
The Lanczos algorithm is stopped after a maximum of 20 steps.
Due to numerical instability, sometimes the product $\beta_i\gamma_i$ of Lanczos coefficients becomes negative.
In this case, we stop the continued fraction just before a spurious negative element
$\beta_i\gamma_i <0$ occurs.

\begin{figure}
    \centering
    \includegraphics[width=\textwidth]{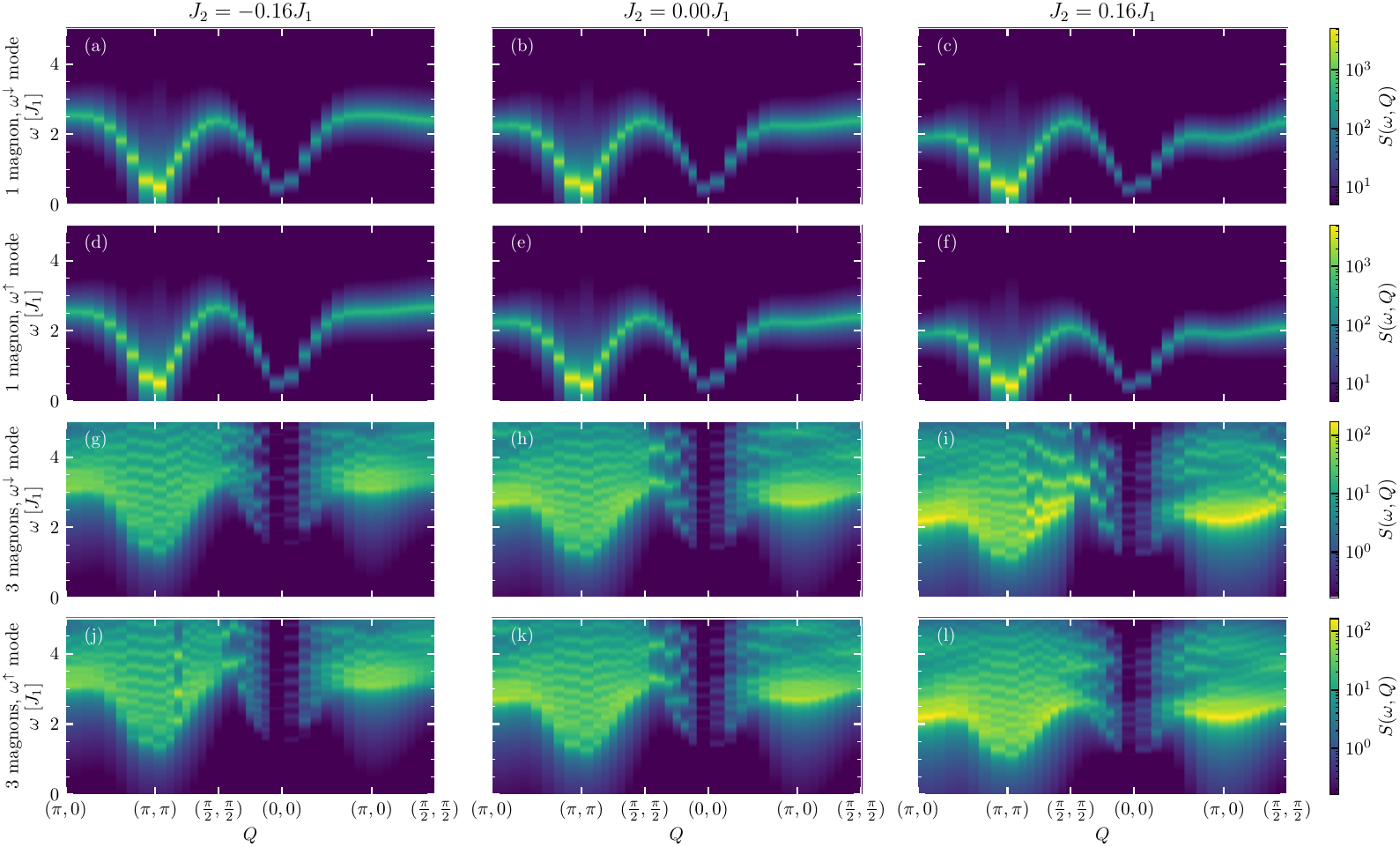}
    \caption{Transversal dynamical structure factors $S^{+-}(\omega,\Vector{Q})$ ($\omega^\downarrow$ mode) and $S^{-+}(\omega,\Vector{Q})$ ($\omega^\uparrow$ mode) separated into their single-magnon [(a)-(f)] and three-magnon contribution [(g)-(l)].
    Note the significantly lower intensity scale for the three-magnon contributions and the logarithmic scale.
    The logarithmic scale has a cutoff at $10^{-3}$ of the maximum value to avoid logarithmic divergence for values near zero.
    }
    \label{results::sd_transverse}
\end{figure}

\begin{figure}
    \centering
    \includegraphics[width=\textwidth]{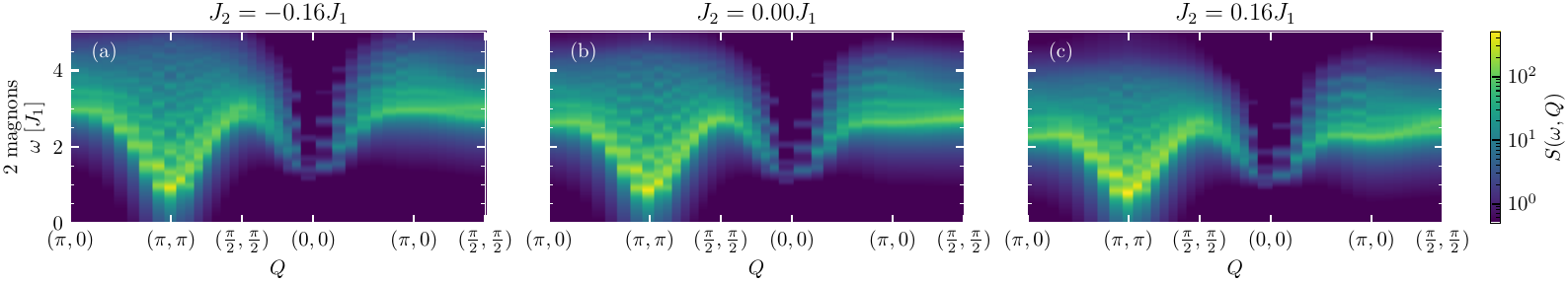}
    \caption{Longitudinal dynamical structure factor $S^{zz}(\omega,\Vector{Q})$ for $J_2=\pm\num{0.16}J_1$ and $J_2 = 0$.
    The logarithmic scale has a cutoff at $10^{-3}$ of the maximum value to avoid logarithmic divergence for values near zero.
    }
    \label{results::sd_longitudinal}
\end{figure}

The resulting spectral densities are shown in \cref{results::sd_transverse} (transversal) and \cref{results::sd_longitudinal} (longitudinal) for the parameters $J_2 = \pm\num{0.16}J_1$ and $J_2 = 0$.
The logarithmic scale for $S(\omega,\Vector{Q})$ is cut off for small values to avoid logarithmic divergence near zero.
In the single-magnon channel, we find sharp magnon bands, which are expected due to the convergence of the \gls{qpc} generator. So, no essential
decay should occur. The discernible width in the panels results from the
artificial broadening necessary for smoothing and plotting.
This demonstrates the approximate stability of the magnon modes within our energy resolution, see App.~\ref{a:connection-between-rod-and-energy-resolution}. 
In the three-magnon channel, a broad continuum with an increased spectral weight at $(\pi,0)$ is observed, with most of the weight at the lower boundaries.
This causes the roton minimum at this $k$-point by level repulsion.
The comparison between the $\omega^\downarrow$ mode and the $\omega^\uparrow$ mode shows the spin splitting, most notably, in the path from $(\pi,0)$ to $(\pi/2,\pi/2)$.
As discussed in Ref.~\cite{liu2024}, we also find an indication for a splitting of the low-energy peak in the two-magnon channel (\cref{results::sd_longitudinal}) at $(\pi/2, \pi/2)$ for $J_2\neq0$.
Note that the area around $\Vector{Q} = (0,0)$ shows almost no weight for the two- and three-magnon channels because for \gls{pbc} the coefficients for $(0,0)$ have been set to zero to avoid a divergence, as mentioned in~\cref{ss:cst}. 
This is no major drawback because rigorously no spectral response is present at the $\Gamma$ point $\Vector{Q} = (0,0)$ because of rotation symmetry in spin isotropic models.

\begin{figure}
    \centering
    \includegraphics[width=\textwidth]{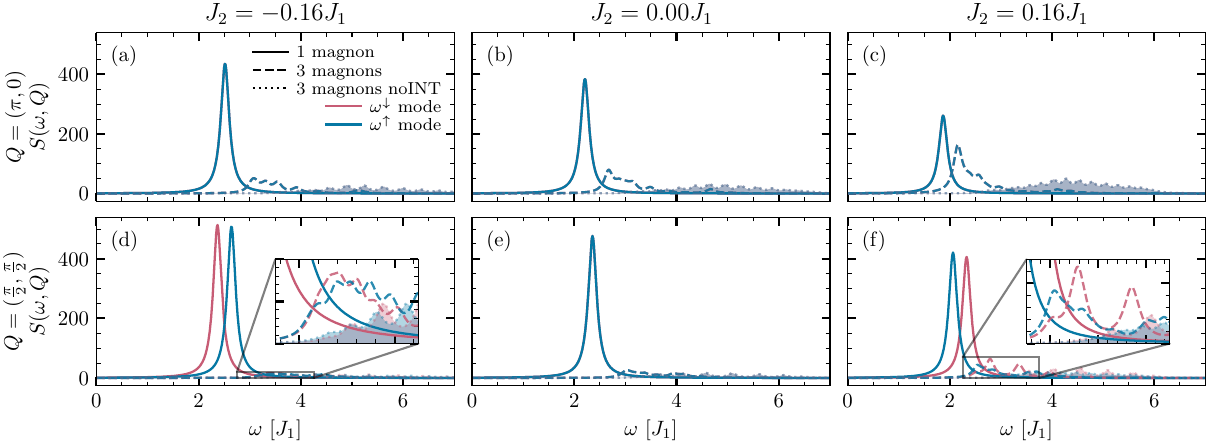}
    \caption{Transversal dynamical structure factor at $\Vector{Q}=(\pi, 0)$ (top row) and $\Vector{Q}=(\pi/2, \pi/2)$ (bottom row) for $J_2=\num{-0.16}J_1$ (left), $J_2=\num{0}$ (middle) and $J_2=\num{0.16}J_1$ (right).
    In addition, the three-magnon contribution without magnon-magnon interaction is shown for both modes by dotted lines and colored shading.}
    \label{results::sd_slices_transverse}
\end{figure}

\begin{figure}
    \centering
    \includegraphics[width=\textwidth]{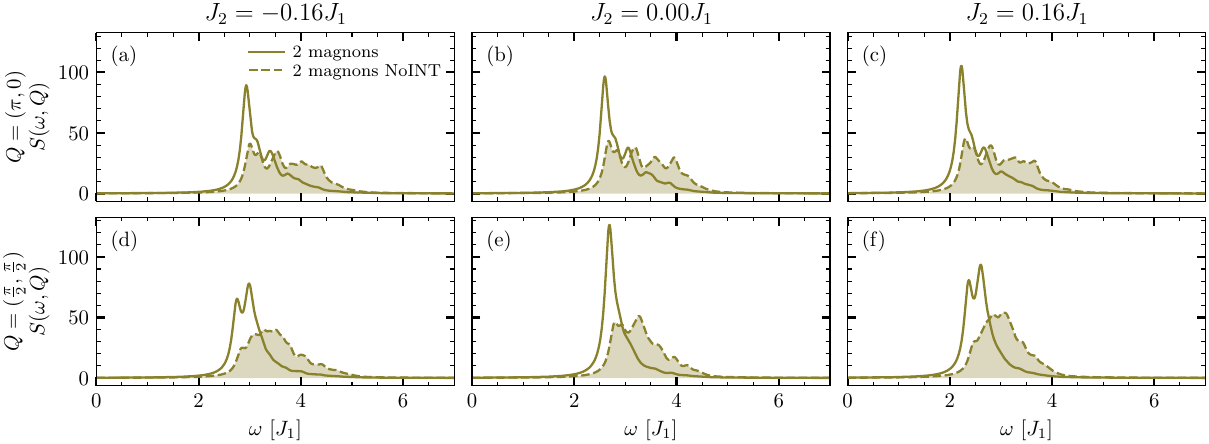}
    \caption{Longitudinal dynamical structure factor with (solid lines) and without magnon-magnon interaction (dashed lines and shading) at $\Vector{Q}=(\pi, 0)$ (top row) and $\Vector{Q}=(\pi/2, \pi/2)$ (bottom row) for $J_2=\num{-0.16}J_1$ (left), $J_2=\num{0}$ (middle) and $J_2=\num{0.16}J_1$ (right).}
    \label{results::sd_slices_longitudinal}
\end{figure}


To see the shifts in energy in more detail, \cref{results::sd_slices_transverse,results::sd_slices_longitudinal} show the spectral densities for two specific points:
$\Vector{Q}=(\pi, 0)$ and $\Vector{Q}=(\pi/2, \pi/2)$. 
At $\Vector{Q}=(\pi, 0)$ in \cref{results::sd_slices_transverse}, no distinction between $\omega^\downarrow$ and $\omega^\uparrow$ mode can be made, since the two modes are degenerate. 
But at $\Vector{Q}=(\pi/2, \pi/2)$, the altermagnetic splitting can be seen 
since the $\omega^\downarrow$ mode lies below the $\omega^\uparrow$ mode for negative $J_2$ and above the $\omega^\uparrow$ mode for positive $J_2$.
This can be observed especially well in the single-magnon contribution.
Additionally, we show the two- and three-magnon contributions resulting from a calculation without magnon-magnon interaction.
Both lines are shifted significantly towards higher energies, similar to the results for the Heisenberg square lattice~\cite{Powalski18}.
This clearly indicates that the magnons interact attractively.
Furthermore, the two-magnon channel (\cref{results::sd_slices_longitudinal}) shows the two distinct peaks at $Q = (\pi/2,\pi/2)$ only in the interacting case.
This indicates that the magnon-magnon interaction is causing this effect.

Considering the shift of the magnon continua when including interactions, we find a notable difference between the cases $J_2<0$ and $J_2>0$. 
Focusing on the transversal dynamical structure factor at $\Vector{Q}=(\pi, 0)$, i.e., the position of the roton minimum, we find that the three-magnon continuum is strongly enhanced in the case of $J_2>0$ and shifted towards the single-magnon peak, i.e., towards lower energies. 
This explains the increase of $\Delta R$ for positive $J_2$ as discussed for the data of \cref{results::roton_depth}.
This finding corroborates that antiferromagnetic \acrlong{nnn} coupling, which competes with the nearest-neighbor one, indeed increases quantum fluctuations and the attractive interaction between magnons in particular. 
In fact, this also causes a sizable reduction in spectral weight of the single-magnon mode, i.e., spectral weight is shifted from the single-magnon peak into the continuum. 

In contrast, for $J_2<0$, the continuum displays a decreased spectral weight as compared to the case of $J_2=0$, and the single-magnon spectral weight is larger
in return. At $\Vector{Q}=(\pi/2, \pi/2)$, i.e., at the point with maximum spin splitting, a similar behavior appears, although one has to consider both magnon modes and their three-magnon continua.

Previous studies calculated transversal and longitudinal dynamical structure factors for magnons in 2D altermagnets, considering magnon-magnon interactions~\cite{GarciaGaitan2025, eto2025spontaneousmagnondecaysnonrelativistic,liu2024}, demonstrating the spin splitting in the spectra. With the \gls{cst} method, we can go beyond this stage and study renormalization and magnon interaction in sectors of different magnon numbers separately. This provides valuable insight into the effects of tuning various parameters
on the spin splitting and the roton minimum,
in particular of tuning the \gls{nnn} coupling inducing altermagnetism.
This underlines that \glspl{cst} are well suited for understanding and predicting \gls{ins} spectra of magnons.
A disadvantage is that this analysis only works well within the parameter range where the magnon blocks do not overlap much. In return, the convergence behavior of 
the \gls{qpc} generator already indicates stability or decay of the
quasi-particle excitations, here magnons. Further research is ongoing to extend
the applicability of \glspl{cst} along the lines indicated in Ref.~\cite{Fischer_2010}.

\FloatBarrier

\section{Conclusions}
\label{s::conclusion}

In this work, we used continuous similarity transformations (\glspl{cst}) in momentum space to study the effects of magnon-magnon interactions on the magnon spectrum of a minimal two-dimensional spin-$1/2$ altermagnetic model.
Recent related studies of this model investigated magnon decay and band renormalization with various methods, such as nonlinear spin-wave theory and
numerical tensor network approaches~\cite{GarciaGaitan2025,eto2025spontaneousmagnondecaysnonrelativistic,costa2024giantspatialanisotropymagnon,Cichutek2025SpontDecay,cichutek2025quantumfluctuationstwodimensionalaltermagnets}.
Our study adds a different perspective on the topic since the \gls{cst} 
allows for a decoupling of the different quasi-particle conserving sectors.
In particular, the employed \gls{qpc} generator only converges if the different sectors do not overlap substantially, while the convergence of the $0n$ generator
corroborates the stability of the ground state with long-range order 
of the \neel{} type.

We used these properties in order to map out the parameter range of the \acrlong{nnn} coupling $J_2$ in which the \gls{cst} with the $0n$ generator converges, indicating the stability of the assumed magnetic order. 
Similarly, we mapped out the region where the \gls{qpc} generator converges, which we interpret as a sign of approximate magnon stability within our numerical accuracy.
Even though strong spontaneous decay of the magnons is not expected in the parameter range of \gls{qpc} convergence, significant quantum effects from magnon-magnon interactions independent from magnon decay are still relevant and impact the magnon bands.
The characteristic feature of altermagnetism in insulators is the non-relativistic spin-splitting of the magnon bands.
On the one hand, we demonstrated that the inclusion of magnon-magnon interactions leads to a reduction of this splitting of $\SI{14}{\percent}$ to $\SI{20}{\percent}$ compared to \gls{lswt}, most notably in the case of competing antiferromagnetic couplings. On the other hand, we emphasize that within the studied parameter range, the spin splitting is a qualitatively robust feature of the magnon bands.

We further studied the roton minimum, a dip in the magnon spectrum at high energies, and its dependence on the strength and sign of the \gls{nnn} coupling.
This feature is a known quantum effect arising from the hybridization of the three-magnon continuum with the single-magnon spectrum in the spin-$1/2$ antiferromagnetic Heisenberg model on the square lattice~\cite{Powalski15}.
We found the roton minimum to emerge also upon inclusion of the \gls{nnn} coupling of different strength and sign.
In particular, strong ferromagnetic coupling stabilizing the \neel{} order slightly suppresses the roton dip while competing antiferromagnetic coupling enhances it.

Finally, we calculated the spectral densities which theoretically describe the outcome of \gls{ins} experiments on magnons. 
Here, using \glspl{cst} includes the discussed interaction effects properly while allowing us to address the spectral densities stemming from the different magnon-number sectors separately.
These calculations revealed important attractive magnon-magnon interactions, shifting spectral weight to lower energies. 
Thereby, the weight in the many-magnon channels is increased while the single-magnon weight is reduced.

In conclusion, we provided enhanced fundamental theoretical understanding of magnon dynamics in a paradigmatic altermagnet. 
These findings will serve as guidelines for future experimental and theoretical investigations in this flourishing field of research.

\backmatter

\section*{Acknowledgements}

We thank Alexander Mook for insightful discussions and gratefully acknowledge financial support by the 
Deutsche Forschungsgemeinschaft (DFG, German Research Foundation) through projects UH 90-14/2 (DBH/GSU) and SCHM 2511/13-2 (MRW/KPS) as well as through Grant No. TRR 360-492547816 (RW).
This research was strongly supported by the provision of HPC resources by the Erlangen National High Performance Computing Center (NHR@FAU) of the Friedrich-Alexander-Universit\"at Erlangen-N\"urnberg (FAU).
KPS acknowledges further financial support by the Munich Quantum Valley, which is supported by the Bavarian state government with funds from the Hightech Agenda Bayern Plus.

\bmhead{Author contributions}
Under the supervision of KPS and GSU, all authors contributed equally to the analysis and interpretation of the data and the preparation of the manuscript.
RW and MRW conceived the main idea.
The program for obtaining the \gls{cst} data is based on previous work and adapted by MRW, VS, and DBH.
All authors reviewed the manuscript.

\bmhead{Data availability}
Data will be published after acceptance on TUDOdata. 

\bmhead{Competing interests}
The authors declare no competing interests.

\begin{appendices}

\section{Self-Consistent Mean-Field Theory for Altermagnetic Interactions}
\label{a:self-consisten-mft-altermagnetic-interaction}

This Appendix contains the derivation of the self-consistent mean-field theory used as a starting point of the \gls{cst}. 
Since the major steps are mostly identical to models already covered by \gls{cst}, see Refs.~\cite{Walther23,Hering24}, only the differences are presented here.
Thus, the application of the Dyson-Maleev transformation, the
mean-field decoupling of real-space expressions, and the Fourier transformation are not further elaborated.
However, after these steps, the resulting Hamiltonian contains relevant deviations.
In the quadratic terms relevant for the Bogoliubov transformation  
\begin{align}
 \mathcal{H}_{\text{MF}}^{\text{quadratic}}   =   g_{\kvec{}}\abos*[\kvec{}] \abos[\kvec{}]  +  f_{-\kvec{}} \bbos*[-\kvec{}] \bbos[-\kvec{}] + h_{\kvec{}} \abos[\kvec{}]\bbos[-\kvec{}] + h_{\kvec{}} \abos*[\kvec{}]\bbos*[-\kvec{}] 
\end{align}
with 
\begin{subequations}
\begin{align}
  \begin{split}
  g_{\kvec{}} &= 
      4 J_1 \left (S -\Delta_{\text{mf}}-n_{\text{mf}}\right)\\
    	&\phantom{=\,}+ 2 J_2 \left(-S + n_{\text{mf}} - t_{\text{mf}} \right)\left(1-\gamma_{a_1}\left(\Vector{k}\right)\right)\\
    &\phantom{=\,}+ 2 J'_2 \left(-S + n_{\text{mf}} - t_{\text{mf}} \right)\left(1-\gamma_{a_2}\left(\Vector{k}\right)\right)
\end{split}\\
  \begin{split}
  f_{-\kvec{}}  &=
      4 J_1 \left (S -\Delta_{\text{mf}}-n_{\text{mf}}\right)\\
    &\phantom{=\,}+ 2 J_2 \left(-S + n_{\text{mf}} - t_{\text{mf}} \right)\left(1-\gamma_{a_2}\left(\Vector{k}\right)\right)\\
    &\phantom{=\,}+ 2 J'_2 \left(-S + n_{\text{mf}} - t_{\text{mf}} \right)\left(1-\gamma_{a_1}\left(\Vector{k}\right)\right)
\end{split}\\
h_{\kvec{}}=h^*_{\kvec{}} &= 4 J_1 \left( S -n_{\text{mf}}-\Delta_{\text{mf}}\right)\gamma_{a_1}\left(\frac{\kvec{}}{2}\right)\gamma_{a_2}\left(\frac{\kvec{}}{2}\right)
\end{align}
\end{subequations}
the terms $g_{\kvec{}}$ and $f_{\kvec{}}$ are no longer equal for $J_2 \neq J'_2$.
As a result, the application of a more generalized Bogoliubov transformation is required.
Here, $n_{\text{mf}}$, $\Delta_{\text{mf}}$, and $t_{\text{mf}}$ are expectation values which stem from the mean-field decoupling and are determined self-consistently together with the solution of the Bogoliubov transformation.
The functions $\gamma_{a_1}(\kvec{})$ and $\gamma_{a_2}(\kvec{})$ arise from the Fourier transformation and are defined as
\begin{align}
    \gamma_{a_1}(\kvec{}) &= \frac{1}{2}\sum_{\Vector{a}\in\{\Vector{a}_1,-\Vector{a}_1\}} \symup{e}^{-\mathrm{i}\Vector{k}\Vector{a}}=\cos\left(\Vector{k}\Vector{a}_1\right)&
    \gamma_{a_2}(\kvec{}) &= \frac{1}{2}\sum_{\Vector{a}\in\{\Vector{a}_2,-\Vector{a}_2\}} \symup{e}^{-\mathrm{i}\Vector{k}\Vector{a}}=\cos\left(\Vector{k}\Vector{a}_2\right)
\end{align}
with the two real-space lattice vectors $\Vector{a}_1$ and $\Vector{a}_2$  
The solution of the generalized Bogoliubov transformation yields two distinct dispersions $\omega_{\alpha/\beta}(\kvec{})$ for the two different flavors of bosons $\albos[]$ and $\bebos[]$.
They can be written as
\begin{align}
        \omega_{\alpha}(\kvec{}) &=\phantom{-}\Delta_{\kvec{}} + \sqrt{R_{\kvec{}}^2- \lvert h_{\kvec{}}\rvert^2}~, &
        \omega_{\beta}(-\kvec{}) &= -\Delta_{\kvec{}} + \sqrt{R_{\kvec{}}^2- \lvert h_{\kvec{}}\rvert^2} \\
     \intertext{with} 
        R_{\kvec{}} &= \frac{g_{\kvec{}}+f_{-\kvec{}}}{2}~,&
        \Delta_{\kvec{}} &= \frac{g_{\kvec{}}-f_{-\kvec{}}}{2}~.
\end{align}
Respectively, the transformation of a single boson is given by
\begin{subequations}
\begin{align}
\abos[\kvec{}]= l_{\kvec{}}  \albos[\kvec{}] + m_{\kvec{}} \bebos[-\kvec{}]\\
\bbos*[-\kvec{}]= m_{\kvec{}}  \albos[\kvec{}] + l_{\kvec{}} \bebos[-\kvec{}]
\end{align}
\end{subequations}
with 
\begin{align}
        &l_{\kvec{}}= - \frac{\mu_{\kvec{}}}{\sqrt{\lvert\mu_{\kvec{}}\rvert^2-\lvert h_{\kvec{}}\rvert^2}}
        \quad \text{,} \quad
        m_{\kvec{}} =     \frac{  h_{\kvec{}}}{\sqrt{\lvert\mu_{\kvec{}}\rvert^2-\lvert h_{\kvec{}}\rvert^2}}
        &\text{and} \quad \mu_{\kvec{}}= \omega_{\alpha}(\kvec{}) + f_{-\kvec{}}~.
\end{align}

With the solution of the Bogoliubov transformation defined, the mean-field parameters can be evaluated self-consistently and are calculated as 
\begin{subequations}
\begin{align}
    n_{\text{mf}} &= \frac{1}{N} \sum_{\Vector{k}} \cexpval{ \abos*[\Vector{k}] \abos[\Vector{k}] } 
            =\frac{1}{N} \sum_{\Vector{k}} m_{\Vector{k}}^2 \label{eqn:n_mf}\\
    \Delta_{\text{mf}} &= \frac{1}{N} \sum_{\Vector{k}} \gamma_{a_1}\left(\frac{\Vector{k}}{2}\right)\gamma_{a_2}\left(\frac{\Vector{k}}{2}\right) \cexpval{ \abos[\Vector{k}] \bbos[\Vector{-k}] } 
            =\frac{1}{N} \sum_{\Vector{k}} \gamma_{a_1}\left(\frac{\Vector{k}}{2}\right)\gamma_{a_2}\left(\frac{\Vector{k}}{2}\right)  l_{\Vector{k}} m_{\Vector{k}} \\
    t_{\text{mf}} &= \frac{1}{N} \sum_{\Vector{k}}\frac{1}{2}\left(\gamma_{a_1}(\Vector{k})+\gamma_{a_2}(\Vector{k})\right) 
            \cexpval{ \abos*[\Vector{k}] \abos[\Vector{k}] } 
        = \frac{1}{N} \sum_{\Vector{k}} \gamma_{a_1}(\Vector{k}) m_{\Vector{k}}^2.
        \end{align}
\end{subequations}
Note that in Ref.~\cite{Walther23}, the corresponding equation of \cref{eqn:n_mf} incorrectly uses $l_{\Vector{k}}^2$ instead of $m_{\Vector{k}}^2$.

After the self-consistent \gls{sct} together with the Bogoliubov transformation, the obtained Hamiltonian serves as the starting point of the \gls{cst}.
It can be divided into different parts with respect to the different number of involved operators 
\begin{align}
    \mathcal{H} =  E_0 + \Gamma + \mathcal{V}
\end{align}
with the ground-state energy $E_0$, all quadratic terms in $\Gamma$ and all quartic terms in $\mathcal{V}$. Furthermore, each term can be split into the already known contribution of the usual Heisenberg model, annotated by $\text{HB}$ and already listed in previous articles~\cite{Powalski18,Walther23}, and the additional \gls{nnn} contributions denoted by $\text{NNN}$
\begin{align}
    \mathcal{H} =  E_0^\text{HB} + \Gamma^\text{HB} + \mathcal{V}^\text{HB}+  E_0^\text{NNN} + \Gamma^\text{NNN} + \mathcal{V}^\text{NNN}~.
\end{align}
Here, only the gls{nnn} contributions are given.
For the Hamiltonian of \cref{eq::altermagnet}, the gls{nnn} contribution to the mean-field ground state energy reads
\begin{align}
\begin{split}
    E_0^\text{NNN} =
     \phantom{+} J_2 N 
    &\left(
     2 n_{\text{mf}}^2
    -4n_{\text{mf}}S
    -4n_{\text{mf}}t_{\text{mf}}
    +2S^2
    +4St_{\text{mf}}
    +2t_{\text{mf}}^2
    \right)\\
    +J'_2 N
    &\left( 
    2n_{\text{mf}}^2
    -4n_{\text{mf}}S
    -4n_{\text{mf}}t_{\text{mf}}
    +2S^2
    +4St_{\text{mf}}
    +2t_{\text{mf}}^2\right)~.
    \end{split}
\end{align}
In Ref.~\cite{Walther23}, a factor of $4$ is 
missing in the equation for $E^{\text{HB}}_0$.
The quadratic part can be subdivided into
\begin{align}
    \Gamma^\text{AM} = \Cata{\kvec1,\kvec2 } \albos*[\kvec1]\albos[\kvec2] + \Cbtb{\kvec1,\kvec2 } \bebos*[\kvec1]\bebos[\kvec2]+
\Cab{\kvec1,\kvec2 } \albos[\kvec1]\bebos[\kvec2] + \Catbt{\kvec1,\kvec2 } \albos*[\kvec1]\bebos*[\kvec2]
\end{align}
with
\begin{subequations}
\begin{align}
\begin{split}
\Cata{\kvec1,\kvec2 } =& 2 \left( n_{\text{mf}} - S - t_{\text{mf}}  \right) \delta \left(\kvec{1}-\kvec{2}\right)  \Big[  \left(J_2+J'_2\right) \big(  l_{\kvec{1}}l_{\kvec{2}} {+} m_{\kvec{1}}m_{\kvec{2}} \big)\\
&- J_2  
    \big( \gamma_{a_1}\left(\kvec{1}\right) l_{\kvec{1}}l_{\kvec{2}} {+}\gamma_{a_2}\left(\kvec{1}\right) m_{\kvec{1}}m_{\kvec{2}} \big) - J'_2
    \big( \gamma_{a_2}\left(\kvec{1}\right) l_{\kvec{1}}l_{\kvec{2}} {+}\gamma_{a_1}\left(\kvec{1}\right) m_{\kvec{1}}m_{\kvec{2}} \big)\Big]
\end{split} \\
\Cbtb{\kvec1,\kvec2 } \stackrel{*}{=}& \Cata{\kvec2,\kvec1}  \Theta(\kvec{1}-\kvec{2}) \quad(*\quad\text{with\quad} \gamma_{a_1} \leftrightarrow \gamma_{a_2}) \\
\begin{split}
\Catbt{\kvec1,\kvec2 } =& 2 \left( n_{\text{mf}} - S - t_{\text{mf}}  \right) \delta \left(\kvec{1}+\kvec{2}\right)  \Big[  \left(J_2+J'_2\right) \big(  l_{\kvec{1}}m_{\kvec{2}} + l_{\kvec{2}}m_{\kvec{1}} \big)\\
&- J_2  
    \big( \gamma_{a_1}\left(\kvec{1}\right) l_{\kvec{1}}m_{\kvec{2}} {+}\gamma_{a_2}\left(\kvec{1}\right) l_{\kvec{2}}m_{\kvec{1}} \big) - J'_2
    \big( \gamma_{a_2}\left(\kvec{1}\right) l_{\kvec{1}}m_{\kvec{2}} {+}\gamma_{a_1}\left(\kvec{1}\right) l_{\kvec{2}}m_{\kvec{1}} \big)\Big]
\end{split}\\ 
\Cab{\kvec1,\kvec2 } \stackrel{*}{=}& \Catbt{\kvec2,\kvec1 } \Theta(-\kvec{1}-\kvec{2}) \quad(*\quad\text{with\quad} \gamma_{a_1} \leftrightarrow \gamma_{a_2}) 
\end{align}
\end{subequations}
Note that all previously and subsequently shown equalities between different coefficients contain a phase factor $\Theta$ that originates from Umklapp processes if all momenta of a coefficient are taken only from the first magnetic Brillouin zone. 
These Umklapp processes can lead to an additional phase factor between the coefficients, here denoted by
\begin{align}
\Theta(\Vector{\Gamma}_{\kvec{}})=\gamma_{a_1}\left(\frac{\Vector{\Gamma}_{\kvec{}}}{2}\right) \gamma_{a_2}\left(\frac{\Vector{\Gamma}_{\kvec{}} }{2}\right)
\end{align}
where $\Vector{\Gamma}_{\kvec{}}$ is a reciprocal lattice vector. 
Therefore, the factor only takes values $\Theta\in\{-1,1\}$. 
The exact form of $\Vector{\Gamma}_{\kvec{}}$ is determined by the coefficient on the left side of the equality. 
More precisely, it is encoded in the superscript coefficient operator, where an operator with momentum
$\kvec{i}$ contributes to $\Vector{\Gamma}_{\kvec{}}$ as $"+\kvec{i}"$ if it is a creation
operator and as $"-\kvec{i}"$ if it is an annihilation operator.
This has to be taken into account if any symmetries between the coefficients are used throughout the calculations. 
In addition, the exchange of $\gamma_{a_1} \leftrightarrow \gamma_{a_2}$ can also be understood as a rotation of \SI{90}{\degree} of all the momenta involved on 
the right-hand side. 

Next, the quartic part can be subdivided in
\begin{align}
   \nonumber  \mathcal{V}^\text{AM} = 
\phantom{+}    \Catabtb{\kvec1,\kvec2,\kvec3,\kvec4  } \albos*[\kvec1]\albos[\kvec2]\bebos*[\kvec3]\bebos[\kvec4] 
                    +\Catataa{\kvec1,\kvec2,\kvec3,\kvec4  } \albos*[\kvec1]\albos*[\kvec2]\albos[\kvec3]\albos[\kvec4] 
                   +\Cbtbtbb{\kvec1,\kvec2,\kvec3,\kvec4  } \bebos*[\kvec1]\bebos*[\kvec2]\bebos[\kvec3]\bebos[\kvec4] \\ \nonumber
                    +\Catatbtbt{\kvec1,\kvec2,\kvec3,\kvec4  }\albos*[\kvec1]\albos*[\kvec2]\bebos*[\kvec3]\bebos*[\kvec4] 
                    +\Catatabt{\kvec1,\kvec2,\kvec3,\kvec4  }\albos*[\kvec1]\albos*[\kvec2]\albos[\kvec3]\bebos*[\kvec4] 
                   +\Cataab{\kvec1,\kvec2,\kvec3,\kvec4  } \albos*[\kvec1]\albos[\kvec2]\albos[\kvec3]\bebos[\kvec4] \\ +\Caabb{\kvec1,\kvec2,\kvec3,\kvec4  } \albos[\kvec1]\albos[\kvec2]\bebos[\kvec3]\bebos[\kvec4] 
                   +\Catbtbtb{\kvec1,\kvec2,\kvec3,\kvec4  } \albos*[\kvec1]\albos*[\kvec2]\albos[\kvec3]\bebos*[\kvec4] 
                   +\Cabtbb{\kvec1,\kvec2,\kvec3,\kvec4  } \albos[\kvec1]\bebos*[\kvec2]\bebos[\kvec3]\bebos[\kvec4] 
 \end{align}
with
\begingroup
\allowdisplaybreaks
\begin{subequations}
    \begin{align}
\begin{split}
\Catabtb{\kvec1,\kvec2,\kvec3,\kvec4} =&\,\frac{2}{N} \delta\left( \kvec{1} - \kvec{2} + \kvec{3} - \kvec{4} \right)  \Big[ \\
        &\,\phantom{+}J_2 \big( \gamma_{a_1}\!\left(\kvec{1}{-}\kvec{2}\right)+\gamma_{a_1}\!\left(\kvec{1}{+}\kvec{3}\right) - \gamma_{a_1}\!\left(\kvec{1}\right) - \gamma_{a_1}\!\left(\kvec{4}\right)\!  \big) l_{\kvec{1}}l_{\kvec{2}}m_{\kvec{3}}m_{\kvec{4}}\\
        &\,{+}J_2 \big( \gamma_{a_2}\!\left(\kvec{1}{-}\kvec{2}\right)+\gamma_{a_2}\!\left(\kvec{1}{+}\kvec{3}\right) - \gamma_{a_2}\!\left(\kvec{1}\right) - \gamma_{a_2}\!\left(\kvec{4}\right)\!  \big) l_{\kvec{3}}l_{\kvec{4}}m_{\kvec{1}}m_{\kvec{2}}\\
        &\,{+}J'_2 \big( \gamma_{a_2}\!\left(\kvec{1}{-}\kvec{2}\right)+\gamma_{a_2}\!\left(\kvec{1}{+}\kvec{3}\right) - \gamma_{a_2}\!\left(\kvec{1}\right) - \gamma_{a_2}\!\left(\kvec{4}\right)\!  \big) l_{\kvec{1}}l_{\kvec{2}}m_{\kvec{3}}m_{\kvec{4}}\\
        &\,{+}J'_2 \big( \gamma_{a_1}\!\left(\kvec{1}{-}\kvec{2}\right)+\gamma_{a_1}\!\left(\kvec{1}{+}\kvec{3}\right) - \gamma_{a_1}\!\left(\kvec{1}\right) - \gamma_{a_1}\!\left(\kvec{4}\right)\!  \big) l_{\kvec{3}}l_{\kvec{4}}m_{\kvec{1}}m_{\kvec{2}}\Big]
\end{split}\\
\begin{split}
\Catataa{\kvec1,\kvec2,\kvec3,\kvec4} =&\, \frac{1}{2N}\delta\left(\kvec{1} + \kvec{2} - \kvec{3} - \kvec{4} \right) \Big[\\
    &\,\phantom{+}J_2\big(
    \gamma_{a_1}\!\left(\kvec{1}{-}\kvec{3}\right) + \gamma_{a_1}\!\left(\kvec{2}{-}\kvec{3}\right) -\gamma_{a_1}\!\left(\kvec{1}\right)- \gamma_{a_1}\!\left(\kvec{2}\right)\!    \big)l_{\kvec{1}}l_{\kvec{2}}l_{\kvec{3}}l_{\kvec{4}}\\
    &\,{+}J_2\big(
    \gamma_{a_2}\!\left(\kvec{1}{-}\kvec{3}\right) + \gamma_{a_2}\!\left(\kvec{2}{-}\kvec{3}\right) - \gamma_{a_2}\!\left(\kvec{1}\right)- \gamma_{a_2}\!\left(\kvec{2}\right)\!    \big) m_{\kvec{1}}m_{\kvec{2}}m_{\kvec{3}}m_{\kvec{4}}\\
        &\,{+}J'_2\big(
    \gamma_{a_2}\!\left(\kvec{1}{-}\kvec{3}\right) + \gamma_{a_2}\!\left(\kvec{2}{-}\kvec{3}\right) -\gamma_{a_2}\!\left(\kvec{1}\right)- \gamma_{a_2}\!\left(\kvec{2}\right)\!    \big)l_{\kvec{1}}l_{\kvec{2}}l_{\kvec{3}}l_{\kvec{4}}\\
    &\,{+}J'_2\big(
    \gamma_{a_1}\!\left(\kvec{1}{-}\kvec{3}\right) + \gamma_{a_1}\!\left(\kvec{2}{-}\kvec{3}\right) - \gamma_{a_1}\!\left(\kvec{1}\right)- \gamma_{a_1}\!\left(\kvec{2}\right)\!    \big) m_{\kvec{1}}m_{\kvec{2}}m_{\kvec{3}}m_{\kvec{4}}\Big]
    \end{split}\\
\Cbtbtbb{\kvec1,\kvec2,\kvec3,\kvec4  } \stackrel{*}{=}& \,
\Catataa{\kvec4,\kvec3,\kvec2,\kvec1} 
\Theta(\kvec{1}+\kvec{2}-\kvec{3}-\kvec{4}) \quad(*\quad\text{with\quad} \gamma_{a_1} \leftrightarrow \gamma_{a_2}) \\
\begin{split}
\Catatbtbt{\kvec1,\kvec2,\kvec3,\kvec4}=&\,\frac{1}{2N}\delta\left(\kvec{1}+\kvec{2}+\kvec{3}+\kvec{4} \right) \Big[\\
&\,\phantom{+}J_2 \big(\gamma_{a_1}\!\left(\kvec{1}{+}\kvec{3}\right) + \gamma_{a_1}\!\left(\kvec{2}{+}\kvec{3}\right) + \gamma_{a_1}\!\left(\kvec{1}\right) + \gamma_{a_1}\!\left(\kvec{2}\right)\!\big)l_{\kvec{1}}l_{\kvec{2}}m_{\kvec{3}}m_{\kvec{4}}\\
&\,{+}J_2 \big(\gamma_{a_2}\!\left(\kvec{1}{+}\kvec{3}\right) + \gamma_{a_2}\!\left(\kvec{2}{+}\kvec{3}\right) + \gamma_{a_2}\!\left(\kvec{1}\right) + \gamma_{a_2}\!\left(\kvec{2}\right)\!\big)l_{\kvec{3}}l_{\kvec{4}}m_{\kvec{1}}m_{\kvec{2}}\\
&\,{+}J'_2 \big(\gamma_{a_2}\!\left(\kvec{1}{+}\kvec{3}\right) + \gamma_{a_2}\!\left(\kvec{2}{+}\kvec{3}\right) + \gamma_{a_2}\!\left(\kvec{1}\right) + \gamma_{a_2}\!\left(\kvec{2}\right)\!\big)l_{\kvec{1}}l_{\kvec{2}}m_{\kvec{3}}m_{\kvec{4}}\\
&\,{+}J'_2 \big(\gamma_{a_1}\!\left(\kvec{1}{+}\kvec{3}\right) + \gamma_{a_1}\!\left(\kvec{2}{+}\kvec{3}\right) + \gamma_{a_1}\!\left(\kvec{1}\right) + \gamma_{a_1}\!\left(\kvec{2}\right)\!\big)l_{\kvec{3}}l_{\kvec{4}}m_{\kvec{1}}m_{\kvec{2}}
\Big]
\end{split}\\ 
\Caabb{\kvec1,\kvec2,\kvec3,\kvec4  } \stackrel{*}{=}&\, \Catatbtbt{\kvec4,\kvec3,\kvec2,\kvec1} 
\Theta(-\kvec{1}-\kvec{2}-\kvec{3}-\kvec{4}) 
\quad(*\quad\text{with\quad} \gamma_{a_1} \leftrightarrow \gamma_{a_2}) \\
\begin{split}
\Catatabt{\kvec1,\kvec2,\kvec3,\kvec4  }=&\, \frac{1}{N} \delta\left(\kvec{1}+\kvec{2}-\kvec{3}+\kvec{4}\right) \Big[\\
&\,\phantom{+}J_2 \big( \gamma_{a_1}\!\left(\kvec{1}{-}\kvec{3}\right)+ \gamma_{a_1}\!\left(\kvec{2}{-}\kvec{3}\right)  - \gamma_{a_1}\!\left(\kvec{1}\right)  - \gamma_{a_1}\!\left(\kvec{2}\right)\! \big) l_{\kvec{1}}l_{\kvec{2}}l_{\kvec{3}}m_{\kvec{4}} \\
&\,        {+}J_2 \big( \gamma_{a_2}\!\left(\kvec{1}{-}\kvec{3}\right)+ \gamma_{a_2}\!\left(\kvec{2}{-}\kvec{3}\right)  - \gamma_{a_2}\!\left(\kvec{1}\right)  - \gamma_{a_2}\!\left(\kvec{2}\right)\! \big) l_{\kvec{4}}m_{\kvec{1}}m_{\kvec{2}}m_{\kvec{3}} \\
&\,        {+}J'_2 \big( \gamma_{a_2}\!\left(\kvec{1}{-}\kvec{3}\right)+ \gamma_{a_2}\!\left(\kvec{2}{-}\kvec{3}\right)  - \gamma_{a_2}\!\left(\kvec{1}\right)  - \gamma_{a_2}\!\left(\kvec{2}\right)\! \big) l_{\kvec{1}}l_{\kvec{2}}l_{\kvec{3}}m_{\kvec{4}} \\
&\,        {+}J'_2 \big( \gamma_{a_1}\!\left(\kvec{1}{-}\kvec{3}\right)+ \gamma_{a_1}\!\left(\kvec{2}{-}\kvec{3}\right)  - \gamma_{a_1}\!\left(\kvec{1}\right)  - \gamma_{a_1}\!\left(\kvec{2}\right)\! \big) l_{\kvec{4}}m_{\kvec{1}}m_{\kvec{2}}m_{\kvec{3}}
\Big]
\end{split}\\
\Cabtbb{\kvec1,\kvec2,\kvec3,\kvec4  } \stackrel{*}{=}&\, \Catatabt{\kvec4,\kvec3,\kvec2,\kvec1} 
\Theta(-\kvec{1}+\kvec{2}-\kvec{3}-\kvec{4}) \quad(*\quad\text{with\quad} \gamma_{a_1} \leftrightarrow \gamma_{a_2}) \\
\begin{split}
\Cataab{\kvec1,\kvec2,\kvec3,\kvec4  } =&\, \frac{1}{N} \delta\left( \kvec{1} -\kvec{2} -\kvec{3} - \kvec{4}\right) \Big[\\
&\,\phantom{+}J_2 \big(\gamma_{a_1}\!\left(\kvec{1}{-}\kvec{2}\right) + \gamma_{a_1}\!\left(\kvec{1}{-}\kvec{3}\right) - \gamma_{a_1}\!\left(\kvec{1}\right) - \gamma_{a_1}\!\left(\kvec{4}\right)\! \big)l_{\kvec{1}}l_{\kvec{2}}l_{\kvec{3}}m_{\kvec{4}} \\
&\,{+}J_2         \big(\gamma_{a_2}\!\left(\kvec{1}{-}\kvec{2}\right) + \gamma_{a_2}\!\left(\kvec{1}{-}\kvec{3}\right) - \gamma_{a_2}\!\left(\kvec{1}\right) - \gamma_{a_2}\!\left(\kvec{4}\right)\! \big)l_{\kvec{4}}m_{\kvec{1}}m_{\kvec{2}}m_{\kvec{3}} \\
&\,{+}J'_2 \big(\gamma_{a_2}\!\left(\kvec{1}{-}\kvec{2}\right) + \gamma_{a_2}\!\left(\kvec{1}{-}\kvec{3}\right) - \gamma_{a_2}\!\left(\kvec{1}\right) - \gamma_{a_2}\!\left(\kvec{4}\right)\! \big)l_{\kvec{1}}l_{\kvec{2}}l_{\kvec{3}}m_{\kvec{4}} \\
&\,{+}J'_2 \big(\gamma_{a_1}\!\left(\kvec{1}{-}\kvec{2}\right) + \gamma_{a_1}\!\left(\kvec{1}{-}\kvec{3}\right) - \gamma_{a_1}\!\left(\kvec{1}\right) - \gamma_{a_1}\!\left(\kvec{4}\right)\! \big)l_{\kvec{4}}m_{\kvec{1}}m_{\kvec{2}}m_{\kvec{3}} \Big]
\end{split}\\
\Catbtbtb{\kvec1,\kvec2,\kvec3,\kvec4  } \stackrel{*}{=}&\, \Cataab{\kvec4,\kvec3,\kvec2,\kvec1} 
\Theta(\kvec{1}+\kvec{2}+\kvec{3}-\kvec{4}) 
\quad(*\quad\text{with\quad} \gamma_{a_1} \leftrightarrow \gamma_{a_2}) ~.
\end{align}
\end{subequations}
\endgroup
We stress that introducing \gls{nnn} interactions does not change the initial form of coefficients for the observables, as they are listed in Ref.~\cite{Powalski18}. 
Moreover, neither the flow equations for all coefficients in the Hamiltonian require further modifications since the \gls{nnn} interaction does not introduce novel operator types.
Only the initial values are different and modified as elaborated above.

\section{Connection between the \gls{rod} and the energy resolution}
\label{a:connection-between-rod-and-energy-resolution}

In this Appendix, we elaborate on the connection between the flow parameter 
$\ell$ and the energy resolution after the flow induced by the \gls{qpc} generator.
This connection constitutes one argument justifying convergence of the \gls{qpc} flow, although single magnons are not expected to be rigorously stable for $J_2\neq0$~\cite{eto2025spontaneousmagnondecaysnonrelativistic}. 

For large values $\ell\to\infty$, it can be shown 
\cite{Mielke_1998,Knetter_2000,dusuel_QuarticOscillatorNonperturbative_2004} 
that off-diagonal elements $h_{nj}$ converge to zero as 
\begin{align}
\label{eq:convergence}
    h_{nj}(\ell\to \infty) \propto \exp\left(-\sign(M(n)-M(j))\left(h_{nn}(\ell\to\infty)-h_{jj}(\ell\to\infty)\right)\ell\right) ,
\end{align}
where $M(n)$ indicates the number of magnons (or quasi-particles in the general case)
that are present in the state $n$, assuming that one uses a basis
where the magnon number is a good quantum number. Clearly, one realizes
that Eq.~\eqref{eq:convergence} implies exponential convergence if the energies
$h_{jj}$ are ordered in the same way as their magnon number $M(j)$, i.e., 
$M(n)> M(j)$ implies $h_{nn}> h_{jj}$ for all $n,j$.

Furthermore, the flow induced by the \gls{qpc} generator has the property of first treating processes with higher energy differences before modifying 
those with lower energy differences~\cite{dusuel_QuarticOscillatorNonperturbative_2004}.  
If the energies are not sorted in ascending order, i.e., $h_{jj}>h_{nn}$ despite
$M(n)> M(j)$, the \gls{qpc} generator is in principle able to re-order
the eigenstates and eigenvalues 
if the flow is computed exactly. But in practice, we always have to rely 
on truncations for evaluating the flow so that errors cannot be avoided, and re-ordering is likely to introduce significant inaccuracies.
In fact, we usually observe that the flow diverges for non-sorted eigenstates since the prefactor of $\ell$ in the argument
of the exponential becomes positive.

\begin{figure}
    \centering
    \includegraphics[width=.7\textwidth]{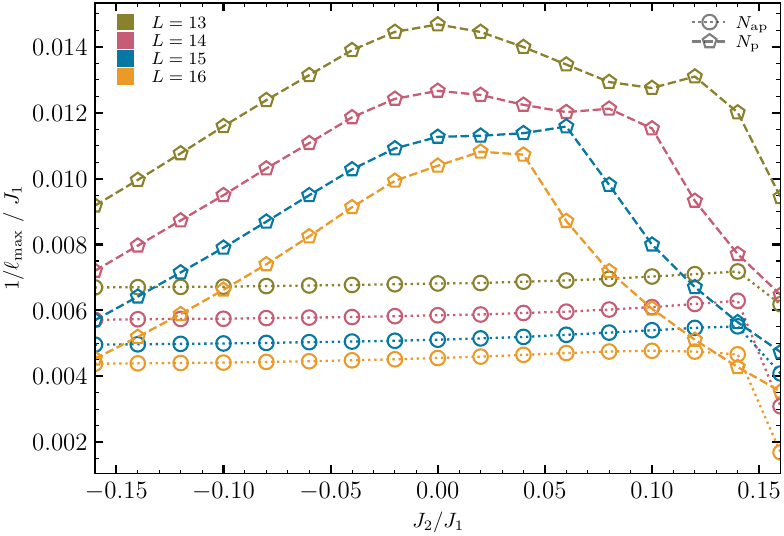}
    \caption{
    Inverse of maximal flow parameter $\ell_{\text{max}}$ where the \gls{rod} undercuts the threshold of $10^{-6} J_1$ depending on $J_2$ with $J'_2=0$.
    Different system sizes $L$ and both boundary conditions 
    \gls{pbc} ($\NP{}$) and \gls{apbc} ($\NAP{}$) are shown.}
    \label{fig::l_max}
\end{figure}

In practice, however, if the negative energy differences are genuinely small compared to other energy scales and occur only for a small number of states, the onset of the corresponding divergence of the flow is not yet noticeable when the total 
\gls{rod} falls below the selected threshold value of $10^{-6}J_1$. 
Then, the inverse of the flow parameter $\ell_\text{max}$ at which the flow undercuts the threshold of the \gls{rod}, provides an estimate of 
the energy resolution that the flow has reached, i.e., how well energetically
different states are separated. 
In \cref{fig::l_max}, the inverse of this maximal flow parameter $\ell_{\text{max}}$ is shown for different system sizes and boundary conditions in the analyzed range of $J_2$ in this work. 
From this plot, one can conclude that decay processes with line widths smaller than $1/\ell_{\text{max}}=0.004J_1$ are not resolved. This means that such
decay processes are essentially ignored, i.e., one does not distinguish
between a $\delta$ distribution and a sharp Lorentzian. 
In this sense, such magnons are interpreted as approximately stable in our calculation. For the analogous discussion of finite-size effects, see
the Introduction in \cref{s::intro}.

\end{appendices}


\bibliography{bibliography}

\end{document}